\documentclass[%
 reprint,
superscriptaddress,
preprintnumbers,
 amsmath,amssymb,
aps,
prb,
floatfix,
showpacs,
showkeys
]{revtex4-2}

\usepackage{graphicx}
\usepackage{bm}
\usepackage{xcolor}
\usepackage{subcaption}
\usepackage{dcolumn}       
\usepackage{multirow}
\usepackage{makecell}
\usepackage{soul}
\usepackage{amssymb}
\usepackage{pifont}
\usepackage{hyperref}

\definecolor{ngreen}{HTML}{808000}

\def  \BiSe      {{Bi$_{\mathrm{2}}$Se$_{\mathrm{3}}$}}    

\begin{document}

\setstcolor{red}

\preprint{APS/123-QED}

\title{Twin domain structure in magnetically doped topological insulators}

\author{Jakub \v{S}ebesta}
\email[Corresponding author: ]{sebesta.j@email.cz} 
\affiliation{Charles University, Faculty of Mathematics and Physics, Department of Condensed Matter Physics, Ke Karlovu 5 121 16 Praha 2, Czech Republic}
\author{Karel Carva}
\affiliation{Charles University, Faculty of Mathematics and Physics, Department of Condensed Matter Physics, Ke Karlovu 5 121 16 Praha 2, Czech Republic}
\author{Dominik Kriegner}
\affiliation{Institute of Physics, Academy of Science of the Czech Republic, Cukrovarnická 10, 162 00 Praha 6, Czech Republic}
\affiliation{Institute of Solid State and Materials Physics, Technical University of Dresden, 01062 Dresden, Germany}
\author{Jan Honolka}
\affiliation{Institute of Physics, Academy of Science of the Czech Republic, Cukrovarnická 10, 162 00 Praha 6, Czech Republic}

\date{\today}

\begin{abstract}
Twin domains are naturally present in the topological insulator \BiSe{} and affect strongly its properties. While studies of its behavior for ideal \BiSe{} structure exist, little is known about their possible interaction with other defects. Extra information are needed especially for the case of artificial perturbation of topological insulator states by magnetic doping, which has attracted a lot of attention recently. Employing ab initio calculations based on layered Green's function formalism, we study the interaction between twin planes in \BiSe{}. We show the  influence of various magnetic and non-magnetic chemical defects on the twin plane formation energy and discuss the related modification of their distribution. Furthermore, we examine the change of dopants' magnetic properties at sites in the vicinity of a twin plane, and the dopants'  preference to occupy such sites. Our results suggest that twin planes repel each other at least over distance of $3-4$~nm. However, in the presence of magnetic Mn and Fe defects a close TP placement is preferred. Furthermore, calculated twin plane formation energies indicate that in this situation their formation becomes suppressed. Finally, we discuss the influence of twin planes on the surface band gap.

\end{abstract}


\maketitle

\section{Introduction}

{One of the most characteristic representatives of topological insulators (TI) are three dimensional time reversal symmetry (TRS) protected TIs~\cite{r16_Bansil_TopologyReview,r09_Xia_DiracCones,r11_Qi_TIandSuperCon,r09_Hsieh_ExpDiracCones}.}
Although the bulk band structure contains a band gap, the surface of such materials hosts unique conductive states, which intersect in the so-called Dirac point possessing a linear dispersion~\cite{r10_Hasan_TIreview,r09_Hsieh_ExpDiracCones,r10_Zhang_QLdependence,r16_Bansil_TopologyReview,r13_Cayssol_DiracEq}. Formation of surface electron states originates from the   strong spin orbit coupling (SOC), which {stems form the occurrence of heavy elements as \textit{e.g.} Bi, Se or Te}
and the presence of TRS. It ensures band crossing at high symmetry points of the Brillouin zone~\cite{book_Bernevig_TI,r15_Ortmann_TI}, while no extra crystal symmetry is required
(compare \textit{e.g.} topological crystal insulators
~\cite{r16_Bansil_TopologyReview,r10_Hasan_TIreview}). Combination of strong SOC and TRS leads to the spin polarization of surface bands~\cite{r10_Hasan_TIreview,r11_Qi_TIandSuperCon,book_Bernevig_TI}. Electrons  {occupying states} in the proximity of the Dirac cone with an opposite momentum possess also opposite spins, so-called spin-momentum locking. It ensures \textit{e.g.}  suppression of back-scattering and related {outstanding} surface transport properties~\cite{r07_Konig_QSHE,r15_Ortmann_TI}.

In real applications the influence of defects  could be important, since they might significantly alter  properties of the ideal matter. 
They could be varied intentionally by chemical doping. In the case of TIs it includes particularly  magnetic doping. It leads to breaking of TRS and therefore it opens a surface gap~\cite{r10_Wray_TI_magperturb,r12_Xu_TImagdop}.
Then, not only does a possible control of the surface conductivity appear due to increased surface scattering, but also it could bring about the occurrence of new unique phenomena e.g. QAHE~\cite{ r13_Chang_QAHE}. 
Besides, native defects occur there, naturally. Their presence is hardly controllable and they might have a significant impact on physical properties~\cite{r16_Carva_BiTe,r19_Maca_CuMnAs} as well.
There could exist several kinds of native defects depending on the actual compound and the growth process. These include  twin planes (TP), which are the  focus of this article.

\begin{figure}[t]
\centering
\includegraphics[width=\columnwidth]{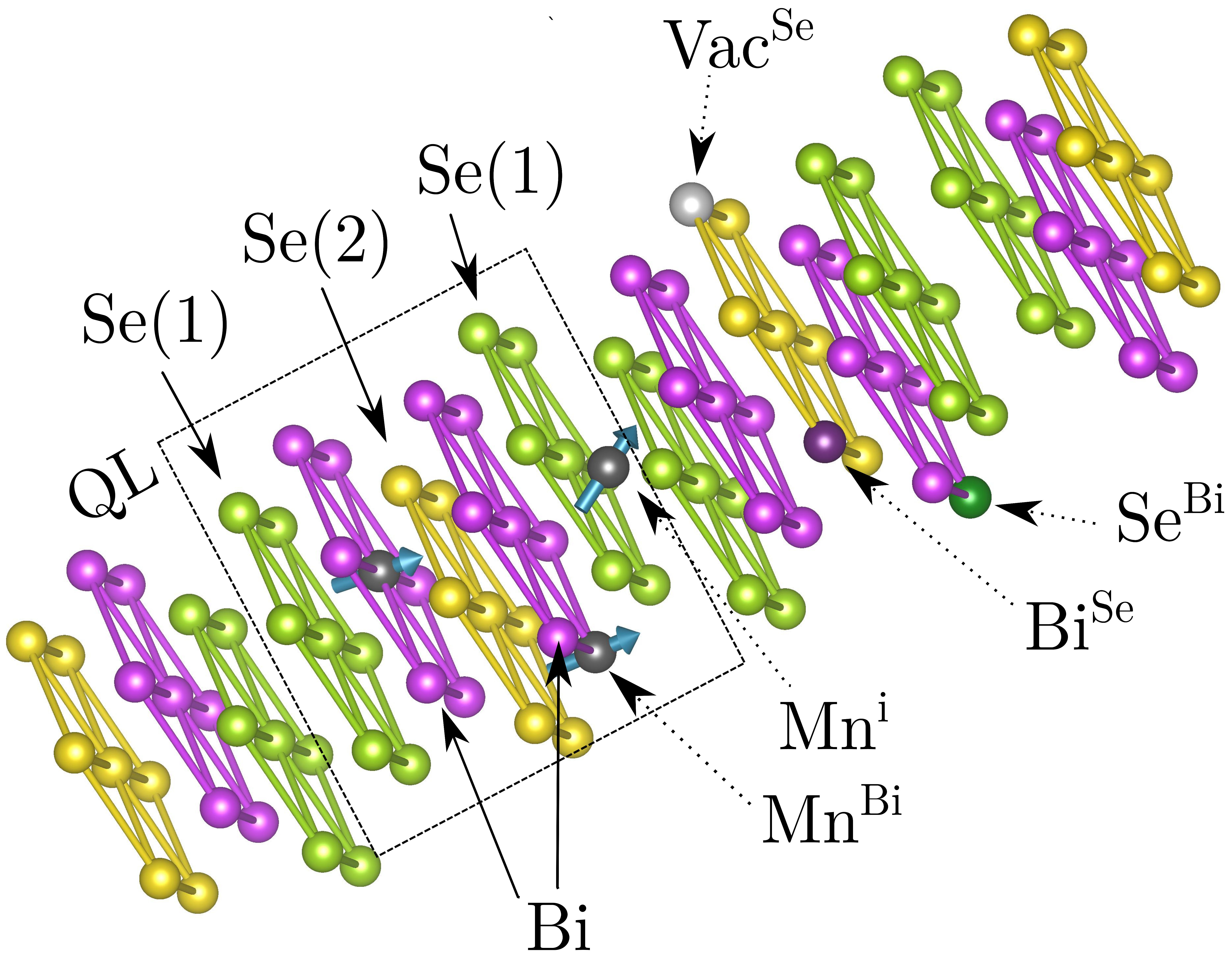}
\caption{Layered crystal structure of \BiSe{}.  Se and Bi layers gathered into QLs are depicted. Examples of (non)-magnetic defects are shown: (Mn$^{\mathrm{Bi}}$) substitutional Mn atoms, (Mn$^{\mathrm{i}}$) interstitial Mn atoms,   (Bi$^{\mathrm{Se}}$) resp. (Se$^{\mathrm{Bi}}$)  Bi and Se anti-sites, (Vac$^{\mathrm{Se}}$) Se vacancies. }
\label{Fig_BiSe}
\end{figure}

An important group of  3D topological insulators are bismuth chalcogenides, such as \BiSe{}~\cite{r09_Zhang_Bi2x3, r10_Zhang_2010} which have been shown to be prone to the formation of twin domains \cite{r14_Lee_BiSe_TP_SeDeficit,r19_Kriegner_TP}.
This compound possesses  a relatively simple band structure, convenient for experimental and theoretical studies, with a Dirac cone appearing at the $\Gamma$ point. The crystal structure of \BiSe{}, belonging to the $R\bar{3}m$ space group, consists of Bi and Se hexagonal layers. 
{They are gathered into quintuple layers (QLs), in which Bi and Se layers alter (Figure~\ref{Fig_BiSe}). Due to  the coupling of QLs only by van der Waals (vdW) forces there  appears a gap between QLs, so-called `van der Waals` gap~\cite{r10_Zhang_2010}}.
The presented crystal structure offers several sites, which could be occupied by magnetic atoms. Based on the theoretical and experimental studies the most probable position occupied by a magnetic dopant  (Cr, Fe, or Mn) is the substitutional position, where  magnetic atoms replace Bi ones ~\cite{r13_Zhang_DopantsPos,r10_Hor_DopantsPosMn,r12_Zhang_dopants}. 
Recently, a formation of septuple layers induced by magnetic defects was described. However, it was shown that it is negligible in  \BiSe{} for small concentrations of magnetic dopants~\cite{r18ArXiv_Rienks_Mnsextuplety}. In our calculation we employ especially Mn$^{\mathrm{Bi}}$ as well as  Fe$^{\mathrm{Bi}}$ magnetic dopants.
Naturally, there also appear native defects like Bi or Se anti-sites (Bi$^{\mathrm{Se}}$ resp. Se$^{\mathrm{Bi}}$), where Bi atoms replace Se ones and \textit{vice versa}. This non-stoichiometry arises due to difficulties in controlling growth conditions, which result in Bi- or Se- rich samples~\cite{r13_Zhang_DopantsPos,r09_Hor_p-type_Bi2Se3_Ca,r12_Scanlon_TIantisites,r16_Wolos_BiSevac,r18_Huang_Bi_antisites}.

\begin{figure}[t]
\centering
\includegraphics[width=\columnwidth]{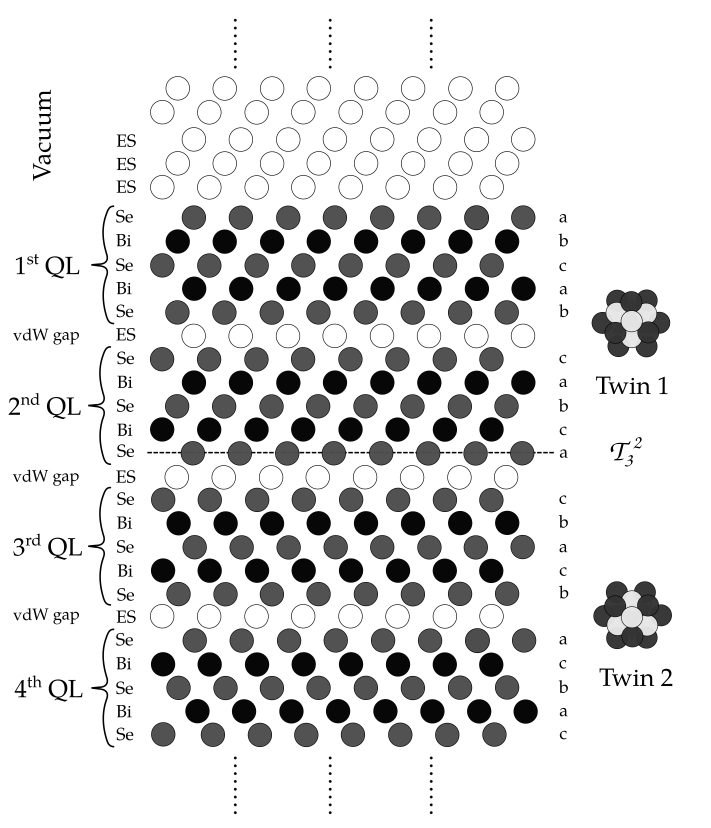}
\caption{Layout of the simulated multilayer \BiSe{} structure including twin planes. Proportions of atoms are neglected.  {QL -- quintuple layer, ES -- empty sphere, $\mathcal{T}^{2}_{3}$ twin plane between the second QL and the third QL.} }
\label{Fig_TP_strukture}
\end{figure}

The above mentioned TPs represent a stacking fault of the layered structure of bismuth chalcogenides~\cite{r12_Medlin_TPaStep,r14_Lee_BiSe_TP_SeDeficit,r19_Kriegner_TP}. Their ideal structure contains three possible stacking positions of hexagonal layers alternating similarly to the  FCC stacking sequence.  During the {formation of the crystal}  there exist two energetically almost equivalent sites, which the atoms in the new layer can choose to occupy. Therefore, mirrored stacking might arise, which could be represented by a 60$^{\circ}$ rotation of new layers in relation to the ideal ones~\cite{r19_Kriegner_TP}. {It results in the inverse order of the \textit{abc}-like stacking beyond the TP (Figure~\ref{Fig_TP_strukture}).} Possibly, there exist a few positions where TP could occur, but the most probable ones lie at outer chalcogenides of quintuples ~\cite{r10_Medlin_TPenergie}. This means that stacking order inside each separate QL contains no defect, the perturbation occurs in the vdW gap between them. QLs after the TP are then constructed with a mirrored stacking order (Figure~\ref{Fig_TP_strukture}). The reported experiments show that the presence of TPs strongly depends on the used substrate~\cite{r14_Tarakina_TPvsSubstrat,r18_Levy_PreGrowProcess}.

Similarly to point defects, TPs might have a significant influence on the physical properties~\cite{r15_Aramberri_TPaDiracSt}.
Therefore, in this paper we focus on the influence of  TPs on 3D TI \BiSe{} behavior and {their interplay with  chemical disorder, especially the magnetic one.
Primarily, we describe a distribution of TPs in a nanoscopically thin \BiSe{} slab. Then, we discuss their  behavior under the presence of magnetic and non-magnetic dopants. Finally, the influence of TPs on the surface states is shown.}

\section{Formalism}

The study employs  ab-initio calculations done in the framework   of  the tight-binding linear muffin-tin  orbital method within the atomic sphere approximation (TB-LMTO-ASA) formulated in terms of Green's functions~\cite{book_Skriver,book_Turek}. It involves the local spin density approximation with the Vosko-Wilk-Nursain exchange-correlation potential~\cite{r80_Vosko_VWNpot} and using of a \textit{s,p,d} atomic model. Calculations were treated in the scalar relativistic approximation, where on-site spin-orbit coupling was involved into the scalar relativistic Hamiltonian as a perturbation. A basic screened impurity model was included to improve treating electrostatics of disordered systems~\cite{r95_Ruban_screen}. Thanks to using of the Green's function formalism  chemical disorder could be included by the coherent potential approximation (CPA)~\cite{r68_Velicky_cpa}. It allows to avoid using of large statistical ensembles and it  is suitable for small perturbation in the system. 
To simulate a layered structure, where it is important to treat TPs, layered Green's functions reflecting translation symmetry only within an atomic layer were {employed}~\cite{book_Turek,r00_Kudrnovsky_layers,r95_Turek_layers,r02_Kudrnovsky_layers}.
In calculations, a multilayer system is attached to the semi-infinite leads, which have to satisfy self-consistent conditions. {Due} to the coupling of the multilayer to the attached leads it is possible to obtain a self-consistent solution also for the inner layers.
{Based on the down-folding method one is able  to construct recursively  embedding potentials acting from both sides on the particular layer, which are related to the interlayer coupling. For a detailed description we refer the reader to Ref.~\cite{book_Turek,r00_Kudrnovsky_layers}.  }

The crystal structure is based on  experimental lattice parameters (\BiSe{} unit cell $a$~=~4.138~{\AA} and $c$~=~28.64~{\AA}~\cite{r10_Zhang_2010}), which were used to build \BiSe{} multilayer structures. The vdW gap in between QLs is included within ASA by placing appropriate empty spheres (ES). {To avoid effects of the substrate (or leads) and to concentrate only on the behavior of proper \BiSe{} layers we  surround it by vacuum, which is treated in a similar sense to the vdW gap}. 
It is formed from the fcc-like stacked empty sphere layers keeping the three-fold symmetry of \BiSe{} layers.
Further, because leads should fit to the simulated structure  slightly modified scandium is selected.  Its hcp crystal structure suits to fcc like stacking within QLs and it possesses not too much distinct lattice parameters~\cite{r56_Spedding_scandium}. {However, leads are more less unimportant  thanks to used vacuum spacers}. 

{Finally, one is able to construct a layer structure, which  consists of intermediate Sc layers at borders, coupled to semiinfinite leads, and several \BiSe{} QLs enclosed by the vacuum spacer. In our calculation we employed ten or twenty QLs wide \BiSe{} structures and the vacuum spacers about ten ES layers width. These  dimensions are  sufficient to simulate the vacuum and to obtain surface gapless states.}  Native defects (Bi$^{\mathrm{Se}}$) as well  as magnetic either Mn$^{\mathrm{Bi}}$ or Fe$^{\mathrm{Bi}}$ doping are included. In general, we assumed homogeneous disorder, where mentioned defects occupy the appropriate sites with the same probability, unless otherwise stated. 
This assumption is supported by synchrotron experiments which show that Mn is not metallic in \BiSe{} and thus does not segregate there~\cite{r16_Valiska_BiSeHeterostructures}.
The influence of the defects on the crystal structure is reflected by local lattice relaxation similar to previous bulk calculation of Bi$_{\mathrm{2}}$Te$_{\mathrm{3}}$ and \BiSe{}~\cite{r16_Carva_BiTe,r20_Carva_BiSe}. However, the relaxation corresponding to the presence of surfaces is not included there. In our calculation we simulate TPs in the vdW gaps with  respect to the required 2D periodicity in the layer. Hence no structure boundaries within a layer, which are related to the presence of TP~\cite{r12_Medlin_TPaStep}, are involved.

The selected approach unfortunately introduces numerical artifacts within the employed  TB-LMTO-ASA framework, which renders  calculations of absolute formation energy of an extra stacking fault not reliable. Therefore, we deal only with relative formation energy considering the same number and similar type of stacking faults, while we focus on various composition and TPs distribution.    Furthermore, the systematic error depends on the distance of the twin plane from the surface as discussed in the Appendix~\ref{Sec.:Append}. The dependence obtained there is thus subtracted from data presented in Results where applicable.  
{Since we deal with various positions of TPs within the multilayer sample we describe their location
by sub- and superscript denoting adjacent QLs for clarity. 
The notation $\mathcal{T}^{x}_{x+1}$ is used for simplicity, where QLs are enumerated from the top interface.}

%

\section{Results and Discussion}

{
Stacking fault energies related to TP formation in ideal \BiSe{} have already been studied elsewhere~\cite{r10_Medlin_TPenergie,r12_Medlin_TPaStep}. Here we focus on their mutual interaction within \BiSe{} slabs
  and subsequently on their interplay with chemical disorder. }

\subsection{Inter twin plane interactions}

\subsubsection{Structure without disorder}

\begin{figure}
\centering
\includegraphics[width=\columnwidth]{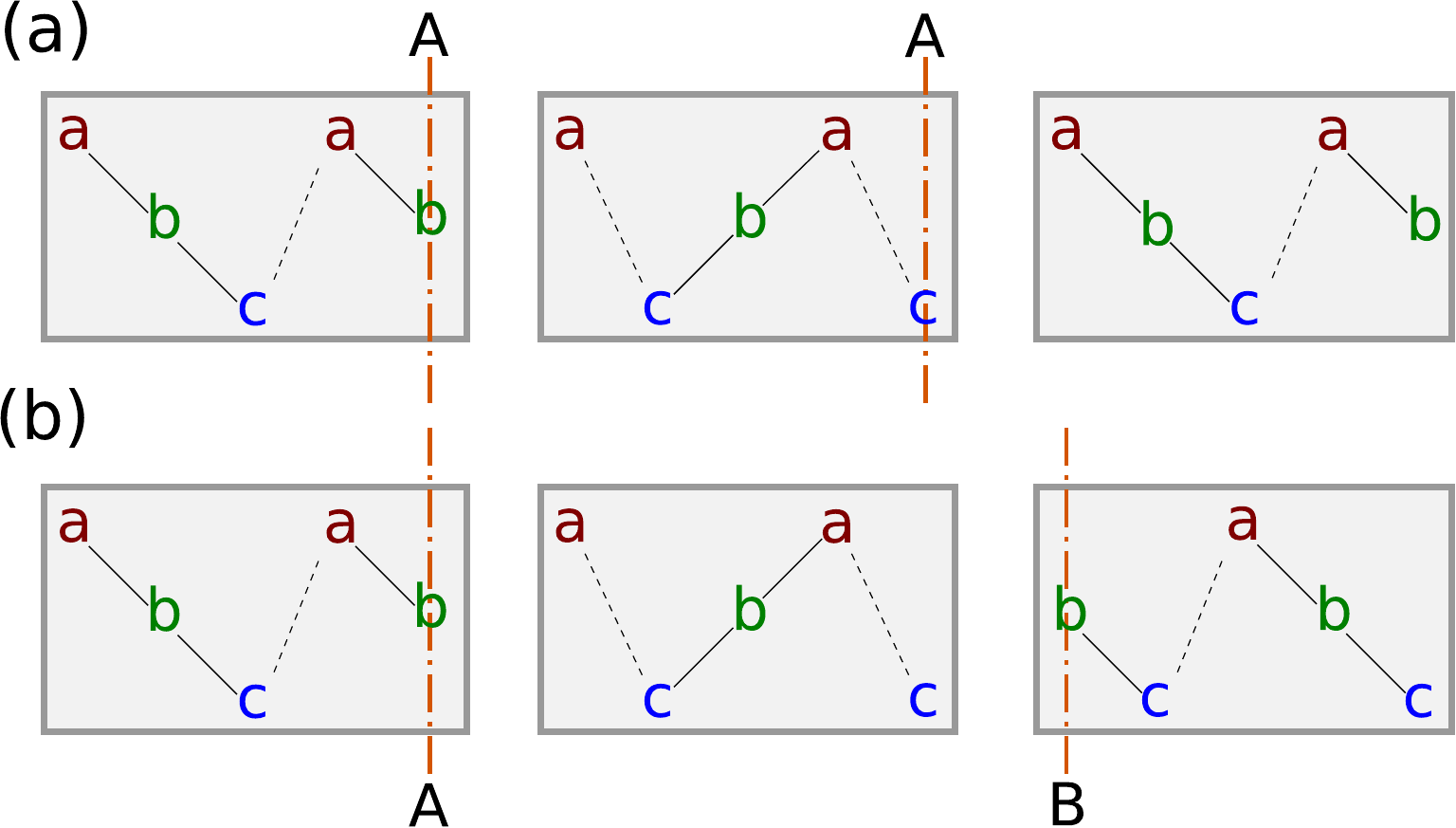}
\caption{Twin plane orientation. There exist two mutual orientation of TPs. The first, TPs are located at the same sides of QLs (similar letters -- AA resp. BB). The second, TPs occurs at the different sides of QLs (distinct letters -- AB).   }
\label{Fig_TP_orientation}  
\end{figure}

\begin{figure}[t]
\centering
\includegraphics[width=\columnwidth]{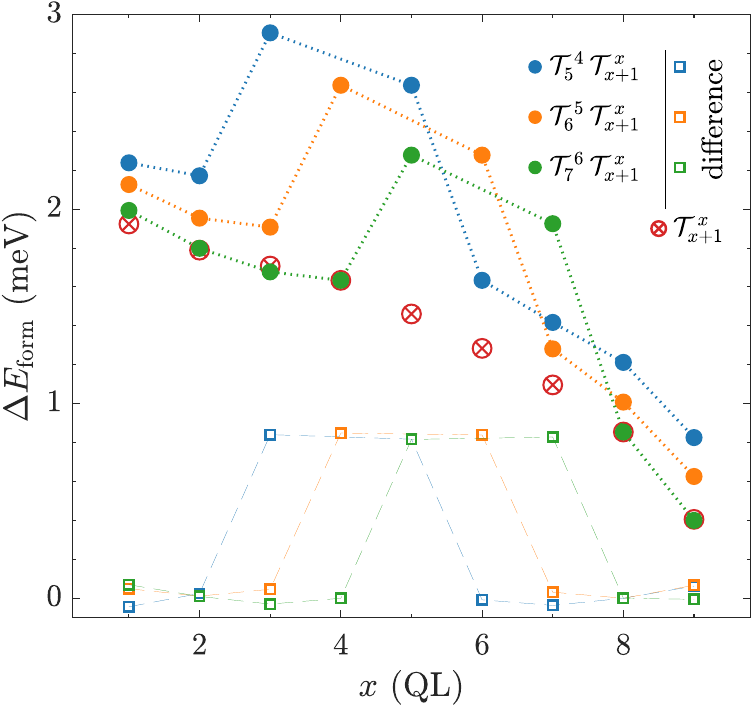}
\caption{(\ding{108}) Relative formation energy of two TPs in dependence on their position within the structure, which consists of 10 \BiSe{} QLs. { ($\otimes$) Relative formation energy of a single TP is depicted for comparison. Relative formation energies belonging to the different numbers of TPs are related to distinct absolute energies. ($\Box$) Energy curves associated to two TPs with subtracted single TP curve contribution.}    }
\label{Fig_2TP_10}
\end{figure}

{One of the simplest ways to study interactions between TPs is the introduction of two TPs  in the pristine \BiSe{} multilayer structure. Keeping the position of one TP fixed while  another one being  independent allows us to determine corresponding relative formation energies over all possible mutual positions of TPs (Figure~\ref{Fig_2TP_10}).}
When studying multiple TPs we have to consider that TPs can occupy the same or distinct sides of particular QLs (Figure~\ref{Fig_TP_orientation}). Different possible cases are compared in Section~\ref{SSec.Inv_TP}. In the remaining text we show for clarity the simplest case with identical orientations of TPs (denoted as  $AA$ according the Figure~\ref{Fig_TP_orientation}, resp. $AAA$ in the case of two extra TPs). 
{ This situation represents qualitatively well   the most probable behavior as it also resembles the lowest energy case of systems with non-identical TPs (Section~\ref{SSec.Inv_TP}).
}

{Considering two TPs in the multilayer structure consisting of 10 QLs, we find that the dependence related to a single TP, discussed in details in the Appendix~(Figure~\ref{Fig_formE}), is changed almost only for  adjacent TPs, where  an extra interaction energy appears~(Figure~\ref{Fig_2TP_10}). {However, for such a small multilayer structure it is not convenient to study  TP interactions because of the strong interplay with the interfaces, which is shown in  the Appendix (Figure~\ref{Fig_formE}a). It is reflected { in bending of calculated energy dependencies (Figure~\ref{Fig_2TP_10}) caused by  non-negligible energy contribution originating from the interactions between TPs and vacuum interfaces (Figure~\ref{Fig_formE})}.}
}

\begin{figure}[t]
\centering
\includegraphics[width=\columnwidth]{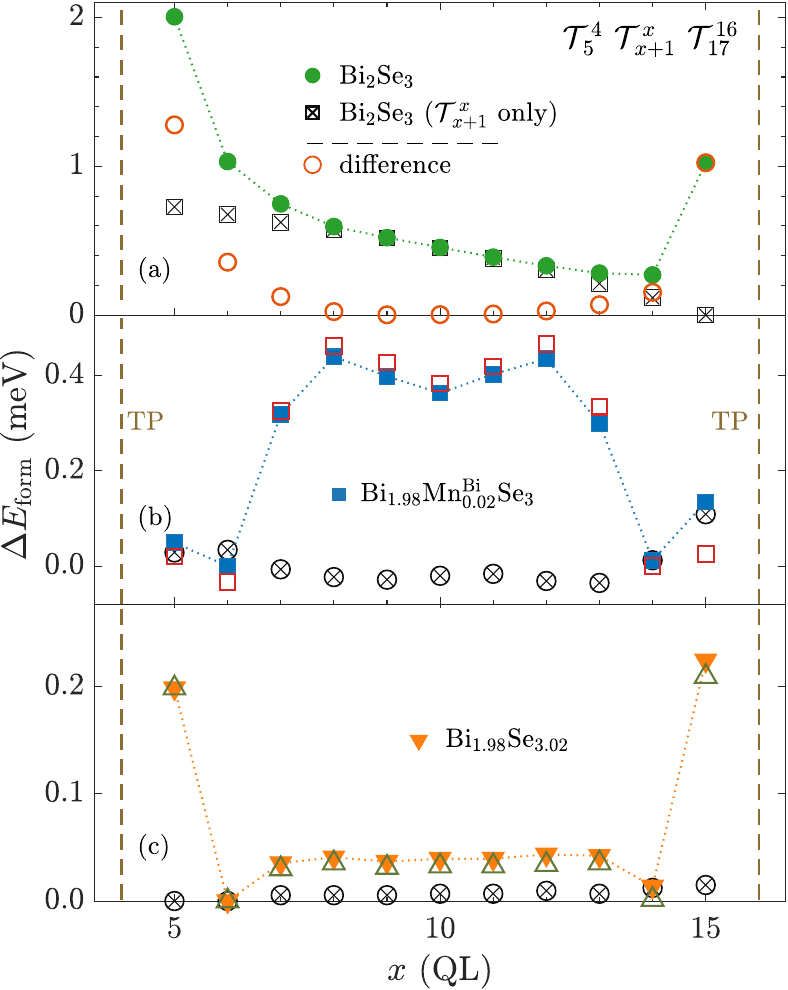}

\caption{(\ding{108},{$\blacksquare$},\ding{116}) Relative formation energy of three TPs as a function of the position $x$  of the middle TP. Positions of border TPs are fixed. The \BiSe{} multilayer consists of 20 QLs.
{(a) pure system without any disorder. (b) system with homogeneous magnetic doping. (c) system under presence of homogeneously distributed nonmagnetic disorder.}
{($\otimes$,$\boxtimes$) Relative formation energy related to a single TP  depicted for comparison.
($\bigcirc$,$\Box$,$\bigtriangleup$) Energy curves associated to three TPs with subtracted single TP curve contribution.}
Particular  relative formation energy curves are  related to different absolute energies.}

\label{Fig_mag_pref}
\end{figure}

Therefore, we introduce a larger structure consisting of 20 \BiSe{} QLs, where positions of two border TPs are fixed   and the third one is able to move in between them~(Figure~\ref{Fig_mag_pref}a). This allows us to study the behavior of a TP in a more realistic situation where it is affected primarily by other surrounding TPs rather than a surface. Interface proximity effects are thus reduced in this situation.. The 3 TP calculation again shows a clearly visible repulsion of neighboring TPs~(Figure~\ref{Fig_mag_pref}a), especially after the subtraction of the surface induced contribution to single TP energy  (Fig~\ref{Fig_formE}).
It reveals the occurrence of a significant interaction energy contribution appearing for TPs distant up to the length of three QLs. This suggests that TPs in a pure sample are likely spread over the sample with mutual distances which exceed at least  the width of three or four QLs. 
Experiments utilizing  X-ray nanobeam microscopy prove that if more TPs are observed, they are clearly several 10~nm apart~\cite{r19_Kriegner_TP}. Hence, one can compare it with the width of one QL, which is about 1~nm~\cite{r10_Zhang_2010}. Nonetheless, one should be aware that we are comparing ground state calculations with a molecular beam epitaxy growth, which occurs far from equilibrium conditions.

\subsubsection{Native and magnetic {defects}}

The interaction between TPs significantly changes when  chemical disorder is introduced in the sample.
For all studied types of doping (Mn$^{\mathrm{Bi}}$ shown in Figure~\ref{Fig_mag_pref}b, or Se$^{\mathrm{Bi}}$ shown in Figure~\ref{Fig_mag_pref}c) we observed a modification of the dependence of the relative formation energy on the distributions of TPs in comparison to the pure sample~(Figure~\ref{Fig_mag_pref}a). 
Thanks to the presence of  disorder the monotonous dependence on the distance from a certain vacuum interface  disappears  (compare Figs.~\ref{Fig_mag_pref}b,c to Figure~\ref{Fig_mag_pref}a; $\otimes$-, $\boxtimes$-points).
{Moreover, the observed relative energy differences are almost one order of magnitude smaller (Figure~\ref{Fig_mag_pref}b,c). It {may be} caused by suppressed interactions between TPs  compared to the ideal case. The presence of a TP  {apparently} does not represent such significant perturbation {in disordered systems} as {it does} in the case of pure, regular systems.}

{Calculations show that {a system with  magnetic disorder}   (Mn$^{\mathrm{Bi}}$) {prefers}  gathering of TPs (Figure~\ref{Fig_mag_pref}b)  instead of their spreading observed in the undoped system. {On the other hand,} non-magnetic disorder rather maintains a repulsion between TPs, although it is quite weak~(Figure~\ref{Fig_mag_pref}c) in comparison to the ideal case~(Figure~\ref{Fig_mag_pref}a), and it {is non-negligible} only for adjacent TPs.} {This  indicates a significance of magnetism related effects, although we cannot exclude the role of different chemistry between dopant species. Therefore we study the influence of the magnetism in more  detail in next sections. }

\subsection{Twin plane formation under chemical disorder}

\begin{figure}
\centering
\includegraphics[width=\columnwidth]{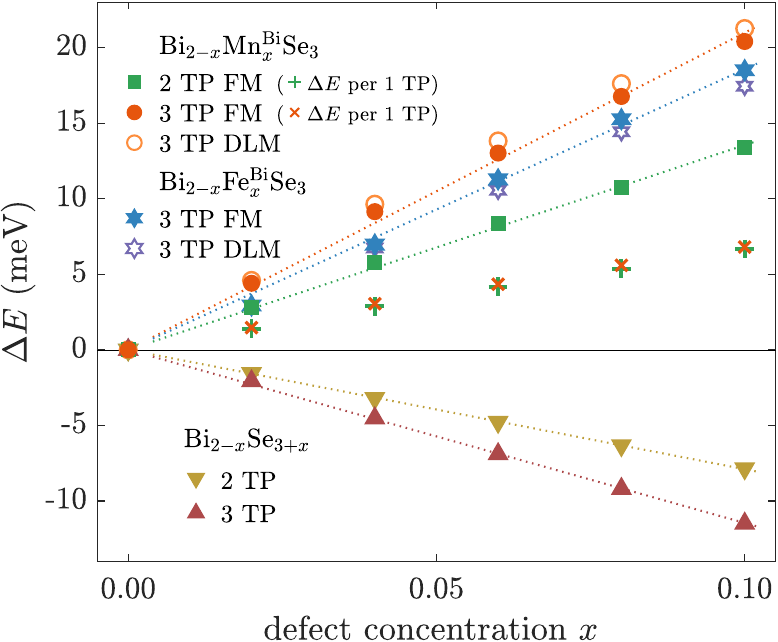}
\caption{ Relative change of the TP formation energy $\Delta E$ in dependence on the concentration of defects $x$. Structures either with two ($\mathcal{T}^{7}_{8}$, $\mathcal{T}^{12}_{13}$ ) or three TPs ($\mathcal{T}^{4}_{5}$, $\mathcal{T}^{9}_{10}$, and $\mathcal{T}^{16}_{17}$ ) within multilayer consisting of 20 QLs were used. Dependencies under presence of magnetic and native defects are depicted. ($+$) and ($\times$) denotes hypothetical relative formation energy related to a single TP. Dotted lines depict calculated linear fits. }
\label{Fig_EnTPinmagdoped}  
\end{figure}

We have calculated dependencies of the relative formation energy of TPs on  the concentration $x$ of magnetic  Mn$^{\mathrm{Bi}}$ or Fe$^{\mathrm{Bi}}$ and native Se$^{\mathrm{Bi}}$ defects to describe the influence of the defect presence on the tendency to TP formation.
(Figure~\ref{Fig_EnTPinmagdoped}). 
The calculated relative formation energy, obtained for different number of included TPs concerning also their distinct positions, almost linearly grows with the increasing concentration of magnetic defects.
It proves that increased amount of magnetic dopants $x$ leads to the suppression of  TPs in the multilayer.  
On the other hand the non-magnetic disorder (Se$^{\mathrm{Bi}}$)  decreases the relative formation energy of TPs in the structure.
However, the appropriate dependencies exhibit linear behavior as well.

{We assume that the suppression of TPs with respect to the increasing concentration of magnetic dopants corresponds to the observed tendency to gathered TPs  in the case of magnetic doping (Figure~\ref{Fig_mag_pref}b). We suppose that gathering of TPs likely minimize an induced effect on the electron structure, which arises from the interplay of TPs and disorder in connection with the magnetism. It agrees with the observation that TPs are less favorable in the magnetically  {doped} systems ~(Figure~\ref{Fig_EnTPinmagdoped}).}
Analogously, the fact that the presence of non-magnetic disorder does favor an occurrence of TPs~(Figure~\ref{Fig_EnTPinmagdoped}) can be related to the suppressed impact of TPs on the system in that case~(Figure~\ref{Fig_mag_pref}c).
Besides, one might note a proportionality of the relative formation energy to the number of the occuring TPs.
It is confirmed by the comparison of the formation energy per single TP (Figure~\ref{Fig_EnTPinmagdoped}).

So far, we discussed ferromagnetically (FM) ordered magnetic dopants. Now, for a moment, we introduce paramagnetic state represented by the  disordered local moment (DLM) model in order to decide whether the TP formation energy depends on the type of the magnetic order.
In general, these two magnetic phases stand for the limiting cases of the magnetic order. One describes a perfectly ordered system, the other one an absolute disorder.
One can observe that calculations exhibit only slight changes of the formation energy with a respect to the former FM order. It indicates that the formation energy likely hardly depends on the type of the magnetic order.
More precisely, TPs become more favorable in the case of Fe doping. 
On the other hand, Mn doping illustrates an opposite behavior.
We suppose that different slopes  of  energy  dependencies induced either by Mn or by  Fe dopants  are likely related to different magnitudes of local exchange splitting. Calculations show that Fe atoms bear about $0.8~\mu_{\mathrm{B}}$ smaller magnitudes of  magnetic moments than the Mn ones. Therefore, one might assume that the TP formation energy likely scales with the size of the change of the  local exchange splitting caused by the mirrored crystal structure.

\begin{figure}
\centering
\includegraphics[width=\columnwidth]{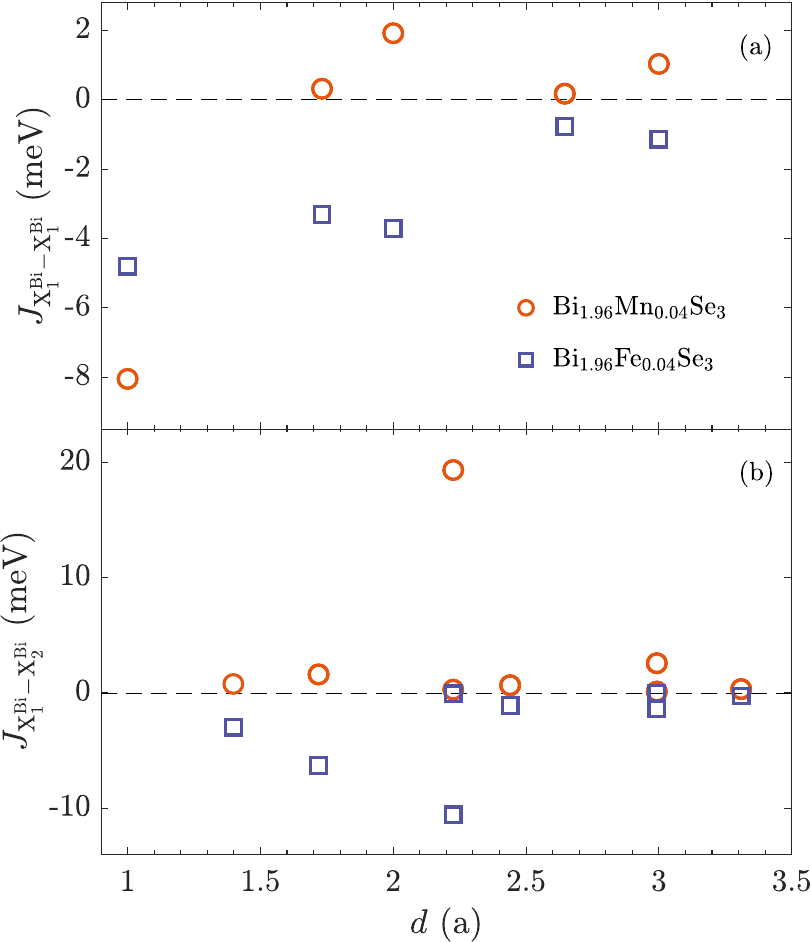}
\caption{ Exchange interaction between magnetic atoms at the substitution position within 
\BiSe{} as a function of the distance in units of the lattice parameter $a$. 
Multilayered structures composed of 20 QLs and FM ordering are employed. Exchange interactions at central QLs are evaluated. 
(a) Exchange interaction within the same sublattice. Interactions within the atomic layer are depicted only in case of layered material. 
(b) Exchange interaction between atoms occupying different sublattices. Interactions over the vdW gap for the layered material are depicted only.}

\label{Fig_ExchInt}  
\end{figure}

The mentioned quite large difference between the magnitudes of Fe and Mn magnetic moment stems from a distinct character of the magnetic exchange interactions (Figure~\ref{Fig_ExchInt}), where they are evaluated by employing the Liechtenstein formula ~\cite{r87_Liechtenstein_exchange,r06_Turek_exchange,r20_Carva_BiSe}.
A comparison shows that unlike Mn related interactions, which are nearly positive except the nearest ones, the exchange interactions between Fe dopants are predominantly antiferromagnetic~\cite{r15_Polyakov_BiFeSe}, regardless of the considered magnetic sublattices.

\subsection{Magnetic dopants behavior}

\begin{figure}
\centering
\includegraphics[width=\columnwidth]{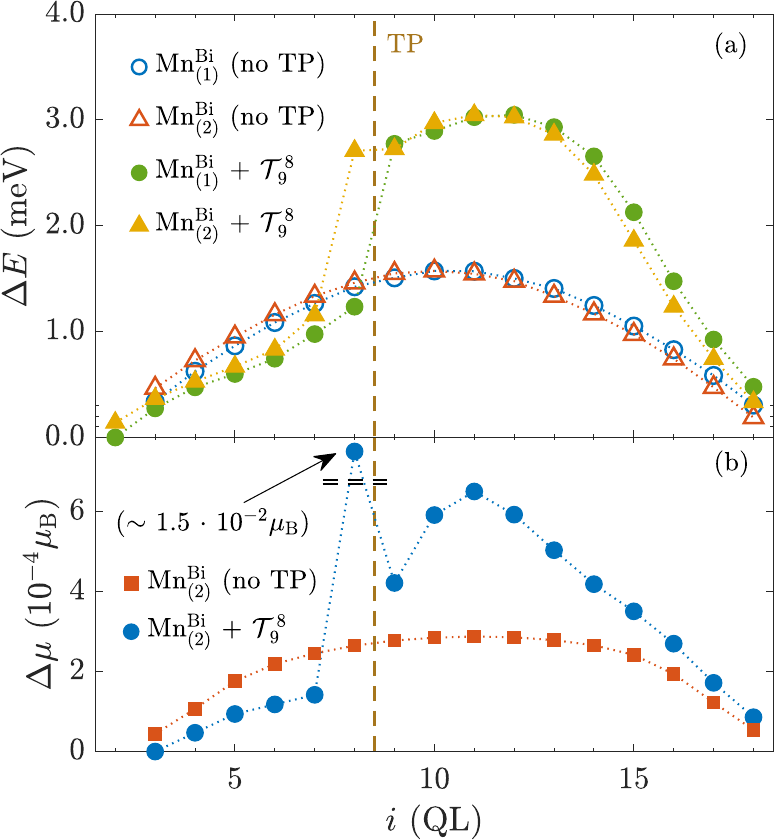}
\caption{Layered magnetic doping of the \BiSe{} multilayer. (a) Relative formation energy of $\mathrm{Mn}^{\mathrm{Bi}}$ substition defects in dependence on the doped QL $i$.
$\mathrm{Mn}^{\mathrm{Bi}}_{{\mathrm{(1)}}}$ and $\mathrm{Mn}^{\mathrm{Bi}}_{{\mathrm{(2)}}}$ stand for substitutions at distinct Bi sites.
$\mathrm{Mn}^{\mathrm{Bi}}_{{\mathrm{(1)}}}$ faces  neighbouring QL with lower $x$.
Dependencies with and without TP are depicted. 
For clarity, they are shifted to fit at the end points.\\
(b) Distribution of magnitudes of $\mathrm{Mn}^{\mathrm{Bi}}$  magnetic moments in dependence of the position of the substitution $i$.  Only one Bi site labeled by the index $i$ of the appropriate QL  is doped by 1\% of Mn.}
\label{Fig_prefdopovani}
\end{figure}

Next, we focused on the influence of TP on the behavior of magnetic dopants, and we calculated relative formation energies of Mn dopants {as a function of}  the position of the dopant determined by indices of the Bi site and QL~(Figure~\ref{Fig_prefdopovani}a). Only one Bi site in the whole structure is partly substituted by Mn.
Comparing the shape of  corresponding curves differing in the presence of a TP, one observes a clear variation of the relative formation energy caused by the TP.
Dependencies without TPs  bear a nearly symmetrical behavior, where a deviation is likely caused by an asymmetry of Bi sites concerning the QL structure. The occurrence of the TP modulates the shape of the former energy dependencies as particular sites become relatively more favored or disfavored according to their location with respect to the TP.  A comparison {of the formation energies~(Figure~\ref{Fig_prefdopovani}a)}   with the magnitudes of induced magnetic moments on Mn dopants~(Figure~\ref{Fig_prefdopovani}b) {indicates a possible} relation between the magnetism or {spin splitting} and the distribution of the relative formation energy of magnetic dopants.
One can observe that the effectively  suppressed relative formation energies, compared to the ideal structure, correspond to the weakening of induced magnetic moments, and vice versa~(Figure~\ref{Fig_prefdopovani}b). The exceptional  {change of the magnitude of the magnetic moment, which is about two orders of magnitude larger than the  other ones,} stems from the proximity of the TP and it can be ascribed to the large charge transfer observed in the undoped structure as described in the Appendix~(Figure~\ref{Fig_charge_transfer}).

\begin{figure}
\centering
\includegraphics[width=\columnwidth]{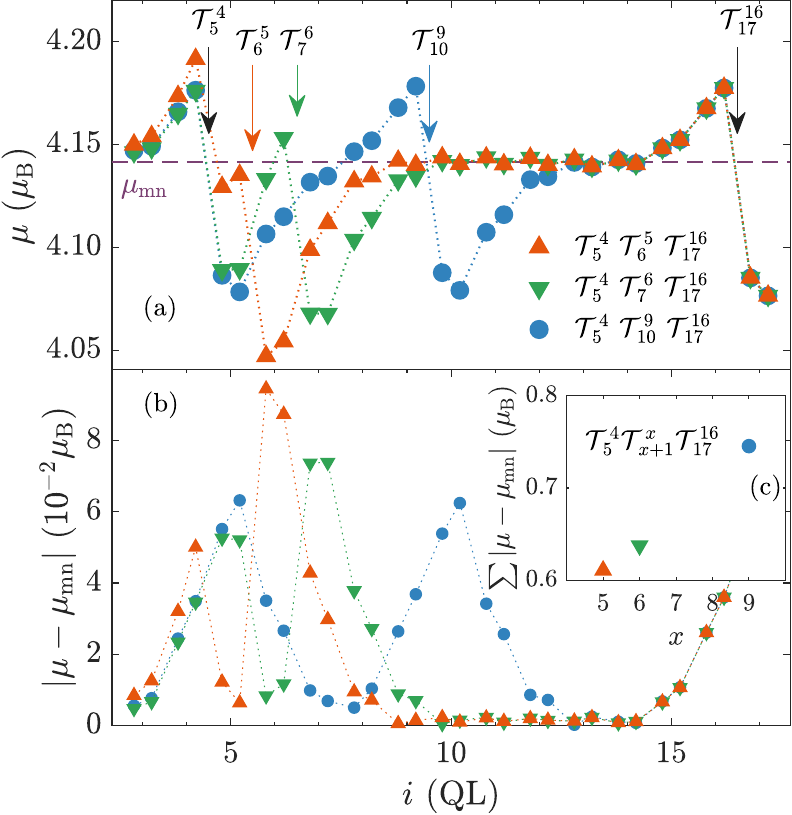}
\caption{(a) Distribution of magnitudes of magnetic moments in homogeneously magnetically  doped Bi$_{\mathrm{1.98}}$Mn$_{\mathrm{0.02}}$Se$_{\mathrm{3}}$ for various positions of TPs. $\mathrm{Mn}^{\mathrm{Bi}}$ are labeled by the index $i$ of the appropriate QL. Arrows point positions of introduced TPs. The outer TPs have fixed position in all the cases (black arrows). Whereas, the inner one is being moved, where the color of the arrow corresponds to the color of the dependence.
(b) {The absolute value of the change of the magnetic moment with the respect to the mean moment value $\mu_{\mathrm{mn}}$. (c) The sum of the magnetic moment changes from the previous subplot.}}
\label{Fig_magmom3TP}
\end{figure}

The described interplay of TPs and magnetic defects could explain the energetic gain observed for gathered TPs in a magnetic material~(Figure~\ref{Fig_mag_pref}). Close TPs lead to a smaller perturbation of the whole electronic structure. This might be deduced from the distribution of calculated magnitudes of magnetic moments in a homogeneously  doped multilayer as a function of the positions of incorporated TPs (Figure~\ref{Fig_magmom3TP}a). We see that the closer TPs are, the smaller is  the {overall} variation of magnitudes of magnetic moments (Figure~\ref{Fig_magmom3TP}c).

\subsection{Comparison of different TP orientations}
\label{SSec.Inv_TP}

\begin{figure}
\centering
\includegraphics[width=\columnwidth]{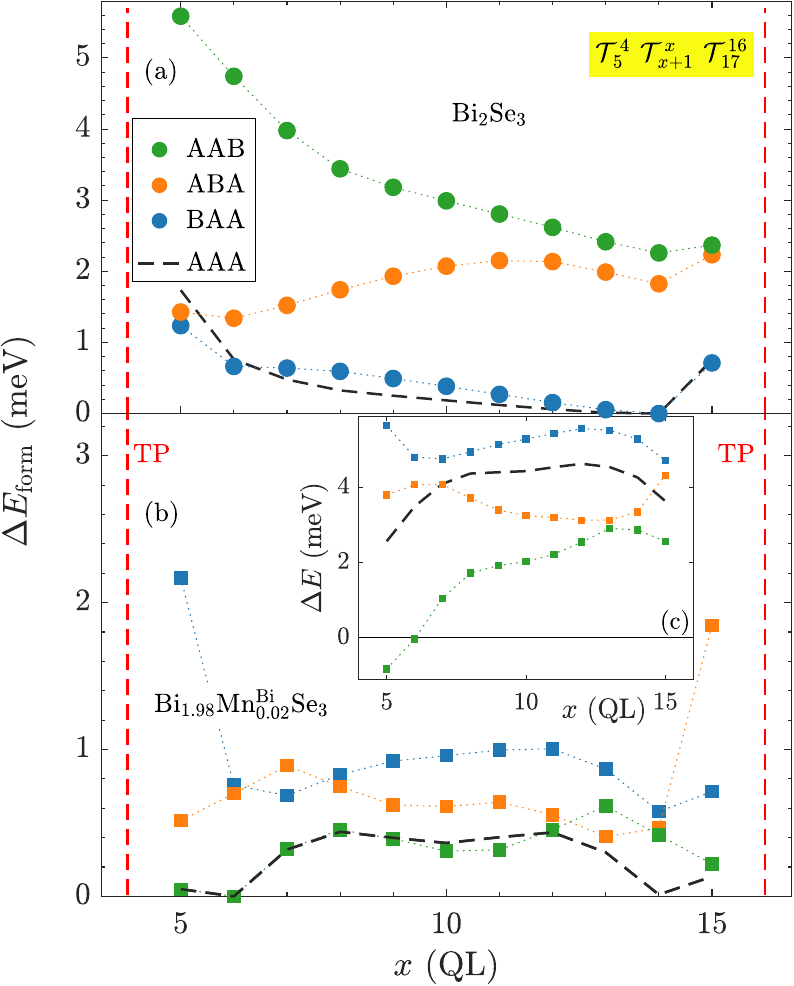}
\caption{Dependence of the relative formation energy of three TPs on the position of the middle TP $x$. Different mutual orientations of the TPs are used. The orientation of TP is labeled by letters A and B. (a) undoped structure. (b) magnetically doped structure. (c) change of the TP formation energy caused by presence of the magnetic defects -- Bi$_{\mathrm{1.98}}$Mn$_{\mathrm{0.02}}$Se$_{\mathrm{3}}$ with respect to the undoped case.
{The relative formation energy curve related to identical TPs ($AAA$ - dashed lines)  depicted in panels (a,b) serves only as a shape reference.}
} 
\label{Fig_TP_form_en_inver}  
\end{figure}

\begin{figure}
\centering
\includegraphics[width=\columnwidth]{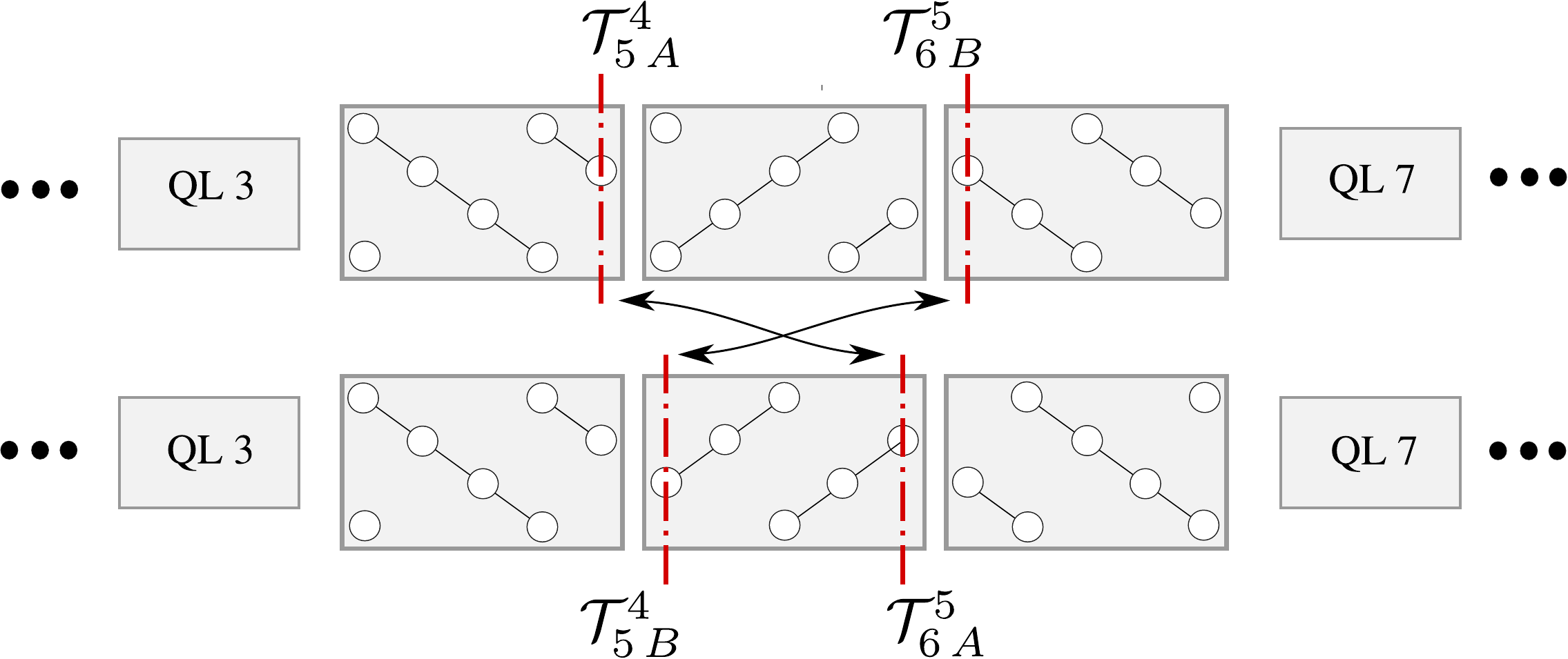}
\caption{Sketch of the interchange of the $AB$ TP order with the $BA$ one in the case of gathered TPs. }
\label{Fig_TPsOrdSwitch}
\end{figure}

In the previous part we described the simplest case consisting in the identical orientations of three TPs ($AAA$)  (Figure~\ref{Fig_TP_orientation}). Now, we focus on the influence of nonidentical TPs on the preceding result. To examine it, we invert the orientation of  each TP in the former three-TP structures, namely the orientations $BAA$, $ABA$, $AAB$  are used (Figure~\ref{Fig_TP_orientation}), and we calculate the distribution of the formation energy.

We recall that in our approach we are not able to compare well the formation energies of the former identically oriented TPs with the case containing a single TP with the inverted orientation. However, the mutual comparison of the new structures is feasible (Figure~\ref{Fig_TP_form_en_inver}). Concerning the undoped structure, one can observe that for the studied number of TPs their formation energy strongly depends  on   the order of TP-type (Figure~\ref{Fig_TP_form_en_inver}a). Considering an increasing index of QLs,  it is evident that the  $BA$ order of two TPs, representing TPs at opposite QL sites (Figure~\ref{Fig_TP_orientation}), is more favorable than the $AB$ one, which stands for TPs at adjacent QL site. Namely, the BAA order, containing no $AB$ sequence,  has the lowest relative formation energy. {Besides, it  is  clearly illustrated at ``touching points'', where two adjacent TPs, either $A$ or $B$ type, are switching (Figure~\ref{Fig_TP_form_en_inver}a),   and the $AB$ order with the $BA$ one are interchanged (Figure~\ref{Fig_TPsOrdSwitch}).  One can assume that  the energy difference  originates from the mentioned  asymmetrical influence of TPs on the surrounding (Figure~\ref{Fig_prefdopovani}), which differentiate the $AB$ and $BA$ order.} {One can notice that the type of the TP sequence can be characterized by number of the vdW gaps in between TPs with the respect to the system of identical TPs. According to the Figure~\ref{Fig_TPsOrdSwitch}), the  $AB$ segment contains an extra vdW gap, whereas the $BA$ segment miss one.    }

{Except the case of the inverted middle TP (ABA), the calculated energy curves {qualitatively} resemble to the relative formation energy curve of identical TPs (Figure~\ref{Fig_TP_form_en_inver}a). Namely, the $BAA$ and $AAA$ are nearly same. The observed disfavor of the $AB$ order
likely gives rise to  a {slightly} higher slope of the $AAB$ energy curve in comparison to the case of identical TPs. The  relative increase of the formation energy might be ascribed to elongation of the segment between $A$- and $B$-typed TPs.  The shape of $ABA$ formation energy curve can be explained in the similar way.}
There occurs a local maximum of the relative formation energy as a function of the position of the middle TPs.{ It likely originates from  a complex interplay arising from the occurrence of two diametrically opposite inter-TP segments  $AB$ and $BA$, while their length is modified.} 

{Considering the symmetry of the \BiSe{} slab placed into the vacuum, where the $BAA$ order is equivalent to the $BBA$ one by a side inversion, one might assume that such immediate alternation of the TP types ($ABA$) is unlikely based on the calculated formation energies. Hence, it appears that the role of the TPs orientation can be regarded as  marginal concerning the distribution of TPs in the undoped structure.}

The relative formation energies dramatically changes under the presence of the magnetic defects similarly to the case of the identical TPs (Figure~\ref{Fig_TP_form_en_inver}b). Although the energy curves are modified by presence of distinct TPs, the local energy minima belonging to gathered TPs are kept. The presence of magnetic dopants reorders the formation energies according to the variation of the TPs orientation. It is likely related to the described interplay of TPs and the magnetic dopands (Figure~\ref{Fig_magmom3TP}). We show that the formation energy of TPs still grows with increased amount of the magnetic dopants nearly irrespective of the TPs positions (compare Figure~\ref{Fig_TP_form_en_inver}c and Figure~\ref{Fig_EnTPinmagdoped}). The existing exceptions originating from the special order of TPs, where TPs are more favorable under magnetic doping (Figure~\ref{Fig_TP_form_en_inver}b). Besides, one should be aware that the calculated curves (Figure~\ref{Fig_TP_form_en_inver}c)  are influenced by vacuum interface induced effects similar to the identical TPs (Figure~\ref{Fig_mag_pref}). Finally, one can conclude that  the TPs orientation does not cause significant qualitative changes in the TPs behavior even in the case of magnetic disorder as the lowest energy case almost mimic the behavior of the system with identical TPs.

\subsection{Surface states}

\begin{figure*}
\centering
\begin{subfigure}{0.25\textwidth}
\includegraphics[width=1.0\textwidth]{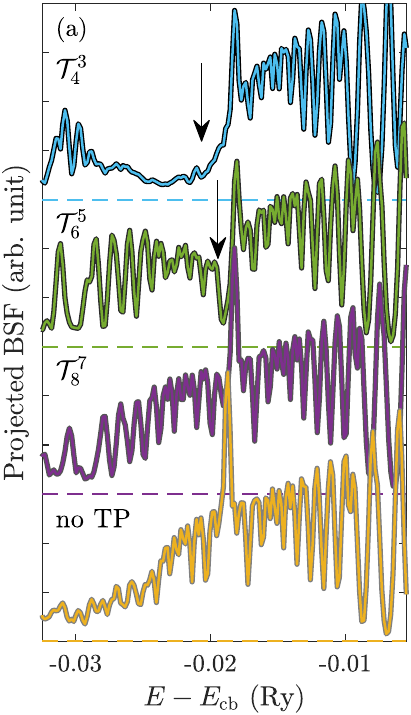}
 \end{subfigure}
\hspace{10pt}
\begin{subfigure}{0.25\textwidth}
\includegraphics[width=\textwidth]{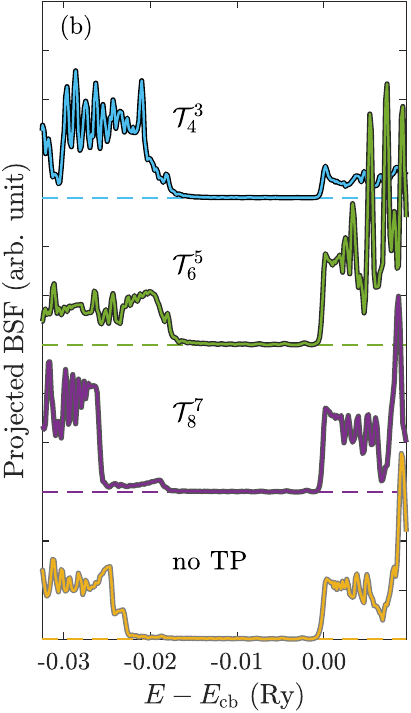}
\end{subfigure}

\caption{Projected BSF of pure \BiSe{} in the vicinity of $\Gamma$ point in dependence on the position of TP. Spin up and spin down channels are overlaying. Obtained surface gaps are denoted by arrows. (a) PBSF of the surface QL , (b) PBSF of the fifth QL from the surface. Energy axis are scaled to the position of conduction band edge~$E_{\mathrm{cb}}$ at the fifth QL. }
\label{Fig_gap_pure}
\end{figure*}

Conductive Dirac surface states are one of the most interesting properties of TIs.. 
The appearance of TPs can strongly influence their presence since the mirroring of the structure symmetry could represent a boundary in the structure. Hence, in this paper we also try  to simulate the influence of the presence of TP and its position on the surface states. We calculated Bloch spectral functions (BSF) in the vicinity of the $\Gamma$ point, where the Dirac cone exists, on the path between high symmetrical reciprocal points $M$ and $K$. In order to study the band gap and surface states we project BSFs it along the mentioned $K-\Gamma-M$ path to the energy-intensity plane in a way that the maximal intensity of the BSF over the $k$-path is selected for particular energy points . Then, a formation of the Dirac states is indicated by vanishing of the energy gap and an occurrence of a strikingly high intensity at the Dirac point, where the surface states intersects~(Figure~\ref{Fig_gap_pure}a).    The {projected} BSFs  (PBSF) reveal that the presence of a TP in at a certain distance from the surface breaks the surface Dirac states, which exist in the unperturbed structure~(Figure~\ref{Fig_gap_pure}a). Our calculations showed that for TPs which are closer than 6 QLs to the surface a gap  opens. In~Figure~\ref{Fig_gap_pure}a it appears especially under presence of $\mathcal{T}_{4}^{3}$.
The oscillations occurring in PBSF dependencies are caused by finite energy- and k-mesh, which prevents obtaining smooth electron bands in terms of the BSF as well as  a narrow k-window, which cuts energy bands. 
Energy scales are related to the position of the well defined conduction band edge~($E_{\mathrm{cb}}$) at an inner QL (Figure~\ref{Fig_gap_pure}b). 
As it was mentioned the observed canceling of surface states and gap opening likely arise from the proximity of two interfaces, which leads to a destructive interference~\cite{r10_Zhang_QLdependence}. Comparing results obtained for TP below the seventh QL with the unperturbed system we found  almost no difference as the Dirac cone is recovered.  Similarly, one can mention a modulation of the bulk band gap width in the vicinity of a TP (Figure~\ref{Fig_gap_pure}b), where a band gap width changes due to the presence of a boundary.  

\section{Conclusions}
We have studied the behavior of TPs in pure layered \BiSe{} system as well as in \BiSe{} under the presence of magnetic and nonmagnetic disorder by first principles calculations.
Calculating total energies of various distributions of several TPs in the multilayer structure, we evaluated the interaction energy between TPs in  dependence on the state of doping. It shows that interaction between TPs in the pure \BiSe{} becomes negligible for distances above three QLs. {However, for smaller distances a significant increase of the TP formation energy was observed, in agreement with the experimentally observed spatial separation of TPs in real samples. The influence of TPs on the surface states of pure \BiSe{} was also studied. We have shown that the surface gap is opened if a TP is closer to it than 6nm.
}

The distribution of TPs and their interplay significantly changes under presence of  chemical disorder. The presence of  non-magnetic disorder weakens the influence of TPs on the electron structure and therefore the interactions between TPs  are significantly smaller. However, the occurrence of  magnetic defects modified the behavior of TPs significantly. Calculations  revealed that adjacent TPs  are energetically more favorable, which corresponds to the dependence of the relative formation energy of TPs on the concentration of magnetic doping.
It reflects a suppression of the TPs formation in  magnetically doped structures, unlike Bi antisites which increase the tendency to TP formation.
{Gathering of TPs leads to a smaller total perturbation of the electron structure and hence might be comprehended as a tendency to TPs annihilation.}
A thorough analysis indicates that the observed mismatch between the magnetic doping and the presence of TPs  consists in  the influence of TPs on the spin splitting of magnetic atoms.

On the other hand, variation of spin splitting, caused by TPs, influence a site preference of magnetic defects within the structure. There occur energetically either more or less preferred sites in the vicinity of the stacking fault, which  can be related to the observed variation of magnitudes of magnetic moments of dopants through the simulated slab. {Mn generally does not prefer to occupy sites right at the twin boundary according to our calculations. Such behavior is indicated also in experiments, since no metallic Mn-Mn bonds were observed while they would probably arise if clustering of Mn at these boundaries was present ~\cite{r16_Valiska_BiSeHeterostructures}. For some sites in the immediate vicinity the energy was lowered, hence the concentration there could be higher than average in the sample.  }

\section*{Acknowledgement}
This research was funded  the Czech Science Foundation (Grant No. 19-13659S). Access to computing and storage facilities owned by parties and projects contributing to the National Grid Infrastructure MetaCentrum provided under the programme ``Projects of Large Research, Development, and Innovations Infrastructures`` (CESNET LM2015042), is greatly appreciated.
This work was supported by The Ministry of Education, Youth and Sports from the Large Infrastructures for Research, Experimental Development and Innovations project ``e-Infrastructure CZ – LM2018140''

\appendix*

\section{Single TP distribution \label{Sec.:Append}}

\begin{figure}[t]
\centering
\includegraphics[width=\columnwidth]{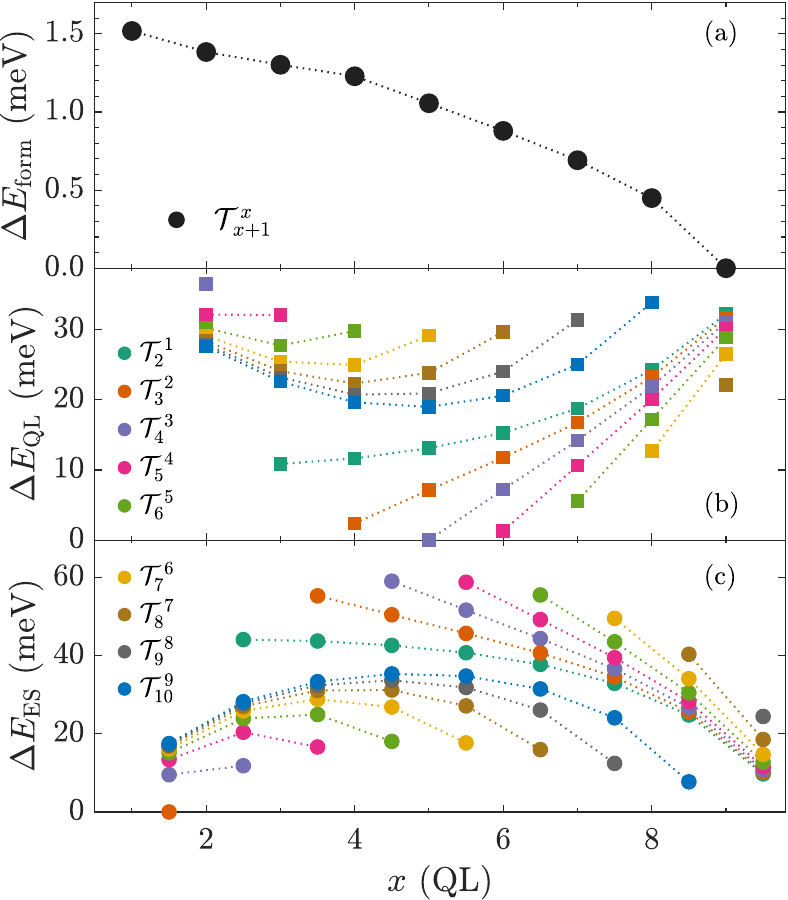}
\caption{(a)  {Single TP relative formation energy $\Delta E_{\mathrm{form}}$} as a function of the position of TP $x$  in the \BiSe{} multilayer.
(b) Relative energy contribution $\Delta E_{\mathrm{QL}}$ of the particular QL $x$. 
(c) Relative energy contribution of ES in dependence on their position.
 \BiSe{} multilayer consists of 10 QLs. Concerning plots (b) and (c), dependencies belonging to several locations of TPs are depicted, where  data points in the vicinity of TPs are excluded for clarity. }

\label{Fig_formE}
\end{figure}

To check behavior of a single TP in a pure multilayer structure and to discuss the vacuum interface caused effect, a distribution of the single TP in a multilayer consisting of 10 QLs is studied.
Thus, an asymmetric dependence of the relative TP formation energy as the function of the TP position was obtained (Figure~\ref{Fig_formE}a).
According to the Figure~\ref{Fig_TP_strukture} a~TP occurs at Se sites in the vicinity of vdW gap.
Generally, one is used to dividing a system to twin domains adjacent to the twin boundary. However, according to the structure composed of QLs, we hypothetically  split the present system into two  domains separated by a vdW gap. It seems to be more convenient to deal with entire QLs in energy comparisons. Nevertheless, one has to be aware that one of domains contains a  mirror layer.
Since we used layer stacking according  Figure~\ref{Fig_TP_strukture}, a TP is located in the domain bearing smaller QL indices.
The relative formation energy  (Figure~\ref{Fig_formE}a) bears a monotonous dependence, which favors a width maximization of  the domain, where the mirror layer belongs. It suggests a significant interplay of the TP with the surface interfaces, which depends on the orientation of the TP.  

\begin{figure}[t]
\centering
\includegraphics[width=\columnwidth]{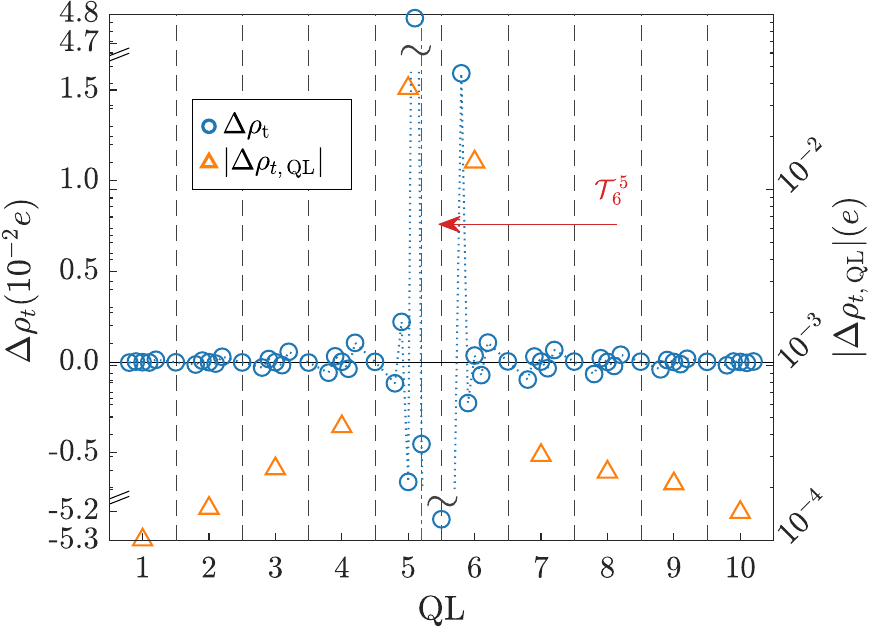}
\caption{ (left axis)
Distribution of the charge transfer $\Delta\rho_{t}$ caused by the presence of a twin plane ($\mathcal{T}^{5}_{6}$) in the multilayer composed of 10 \BiSe{} QLs. Each data point denotes a particular atomic layer or an ES representing the vdW gap. (right axis) {Dependence of the variance of the charge transfer at  particular QLs.} Dashed lines separate QLs and denote position of the vdW gap.}
\label{Fig_charge_transfer}
\end{figure}

Naturally, the occurrence of a TP in the structure causes a charge transfer  compared to the unperturbed system. Calculations show (Figure~\ref{Fig_charge_transfer}) that the electron density changes primarily in the QL possessing a mirror plane of the \BiSe{} layer stacking. Especially, the charge is depleted from the vdW gap and it flows to the Bi layer adjacent to the TP located at Se sites. Besides, a TP causes charge oscillations, which spread out of the TP. Evaluating  {the absolute charge transfer} $\sum \vert \Delta \rho_{t} \vert^{2}$ per a domain  based on the charge transfer distribution  belonging to the system with symmetrical  domain sizes (Figure~\ref{Fig_charge_transfer}), one finds that for the present way of stacking (Figure~\ref{Fig_TP_strukture}) the charge modulation is larger for the domain consisting of QLs with smaller indices. The difference is about one order of magnitude, which implies  a different impact on domains.
Having considered also vacuum interfaces, which represent another structure fault, the system naturally should tend to a suppression of the energetically more demanding perturbations by their separation (Figure~\ref{Fig_formE}a).

Let consider QLs and ES's separately and evaluate their relative total energy contributions. One observes that there occurs a  clear discrepancy between the two presented  domains regardless of the TP position.
Relative magnitudes of total energies $\Delta E_{\mathrm{QL}}$ related to particular QLs $x$ as a function of the position  of the TP
display that the domain containing the TP, namely QLs with the smaller indices,  possesses higher $\Delta E_{\mathrm{QL}}$ and also the closer a TP  to vacuum is the higher the magnitude of $\Delta E_{\mathrm{QL}}$ is. For brevity, we exclude QLs in the vicinity of the TP, since their relative energy changes are of a different scale.
Similarly, relative energy contributions of ES's ($\Delta E_{\mathrm{ES}}$) are also influenced by the position  of the TP (Figure~\ref{Fig_formE}c). However, they follow an opposite evolution compared to the energy contribution of QLs likely because of an opposite impact of the stacking fault. The charge transfer of the related QLs and ES's differs in the sign.
The formation energy curve (Figure~\ref{Fig_formE}a) includes both contributions, $\Delta E_{\mathrm{QL}}$ and $\Delta E_{\mathrm{ES}}$. However, we observe that the shape of the formation energy curve  $\Delta E_{\mathrm{form}}$ (Figure~\ref{Fig_formE}a) is mostly determined by ES's contribution $\Delta E_{\mathrm{ES}}$(Figure~\ref{Fig_formE}c)

It is worthy to study the influence of a single TP on surrounding atomic layers in a pristine structure  as it describes a bare effect of a TP. It shows that the perturbation caused by a TP is hardly local. Dependencies of the charge transfer (Figure~\ref{Fig_charge_transfer}) or the QL resolved total energy (Figure~\ref{Fig_formE}c) indicate that the charge as well as energy modulation spread over few QL. Moreover, a strong interplay with the  \BiSe{} boundaries is visible as it likely yields the shape of the $E_{\mathrm{form}}$ dependence (Figure~\ref{Fig_formE}a).

\bibliography{main}

\providecommand{\noopsort}[1]{}\providecommand{\singleletter}[1]{#1}%
\begin{thebibliography}{46}%
\makeatletter
\providecommand \@ifxundefined [1]{%
 \@ifx{#1\undefined}
}%
\providecommand \@ifnum [1]{%
 \ifnum #1\expandafter \@firstoftwo
 \else \expandafter \@secondoftwo
 \fi
}%
\providecommand \@ifx [1]{%
 \ifx #1\expandafter \@firstoftwo
 \else \expandafter \@secondoftwo
 \fi
}%
\providecommand \natexlab [1]{#1}%
\providecommand \enquote  [1]{``#1''}%
\providecommand \bibnamefont  [1]{#1}%
\providecommand \bibfnamefont [1]{#1}%
\providecommand \citenamefont [1]{#1}%
\providecommand \href@noop [0]{\@secondoftwo}%
\providecommand \href [0]{\begingroup \@sanitize@url \@href}%
\providecommand \@href[1]{\@@startlink{#1}\@@href}%
\providecommand \@@href[1]{\endgroup#1\@@endlink}%
\providecommand \@sanitize@url [0]{\catcode `\\12\catcode `\$12\catcode
  `\&12\catcode `\#12\catcode `\^12\catcode `\_12\catcode `\%12\relax}%
\providecommand \@@startlink[1]{}%
\providecommand \@@endlink[0]{}%
\providecommand \url  [0]{\begingroup\@sanitize@url \@url }%
\providecommand \@url [1]{\endgroup\@href {#1}{\urlprefix }}%
\providecommand \urlprefix  [0]{URL }%
\providecommand \Eprint [0]{\href }%
\providecommand \doibase [0]{https://doi.org/}%
\providecommand \selectlanguage [0]{\@gobble}%
\providecommand \bibinfo  [0]{\@secondoftwo}%
\providecommand \bibfield  [0]{\@secondoftwo}%
\providecommand \translation [1]{[#1]}%
\providecommand \BibitemOpen [0]{}%
\providecommand \bibitemStop [0]{}%
\providecommand \bibitemNoStop [0]{.\EOS\space}%
\providecommand \EOS [0]{\spacefactor3000\relax}%
\providecommand \BibitemShut  [1]{\csname bibitem#1\endcsname}%
\let\auto@bib@innerbib\@empty
\bibitem [{\citenamefont {Bansil}\ \emph {et~al.}(2016)\citenamefont {Bansil},
  \citenamefont {Lin},\ and\ \citenamefont {Das}}]{r16_Bansil_TopologyReview}%
  \BibitemOpen
  \bibfield  {author} {\bibinfo {author} {\bibfnamefont {A.}~\bibnamefont
  {Bansil}}, \bibinfo {author} {\bibfnamefont {H.}~\bibnamefont {Lin}},\ and\
  \bibinfo {author} {\bibfnamefont {T.}~\bibnamefont {Das}},\ }\bibfield
  {title} {\bibinfo {title} {Colloquium: Topological band theory},\ }\href
  {https://doi.org/10.1103/RevModPhys.88.021004} {\bibfield  {journal}
  {\bibinfo  {journal} {Rev. Mod. Phys.}\ }\textbf {\bibinfo {volume} {88}},\
  \bibinfo {pages} {021004} (\bibinfo {year} {2016})}\BibitemShut {NoStop}%
\bibitem [{\citenamefont {Xia}\ \emph {et~al.}(2009)\citenamefont {Xia},
  \citenamefont {Qian}, \citenamefont {Hsieh}, \citenamefont {Wray},
  \citenamefont {Pal}, \citenamefont {Lin}, \citenamefont {Bansil},
  \citenamefont {Grauer}, \citenamefont {Hor}, \citenamefont {Cava},\ and\
  \citenamefont {Hasan}}]{r09_Xia_DiracCones}%
  \BibitemOpen
  \bibfield  {author} {\bibinfo {author} {\bibfnamefont {Y.}~\bibnamefont
  {Xia}}, \bibinfo {author} {\bibfnamefont {D.}~\bibnamefont {Qian}}, \bibinfo
  {author} {\bibfnamefont {D.}~\bibnamefont {Hsieh}}, \bibinfo {author}
  {\bibfnamefont {L.}~\bibnamefont {Wray}}, \bibinfo {author} {\bibfnamefont
  {A.}~\bibnamefont {Pal}}, \bibinfo {author} {\bibfnamefont {H.}~\bibnamefont
  {Lin}}, \bibinfo {author} {\bibfnamefont {A.}~\bibnamefont {Bansil}},
  \bibinfo {author} {\bibfnamefont {D.}~\bibnamefont {Grauer}}, \bibinfo
  {author} {\bibfnamefont {Y.~S.}\ \bibnamefont {Hor}}, \bibinfo {author}
  {\bibfnamefont {R.~J.}\ \bibnamefont {Cava}},\ and\ \bibinfo {author}
  {\bibfnamefont {M.~Z.}\ \bibnamefont {Hasan}},\ }\bibfield  {title} {\bibinfo
  {title} {Observation of a large-gap topological-insulator class with a single
  dirac cone on the surface},\ }\href {https://doi.org/10.1038/nphys1274}
  {\bibfield  {journal} {\bibinfo  {journal} {Nature Physics}\ }\textbf
  {\bibinfo {volume} {5}},\ \bibinfo {pages} {398 EP } (\bibinfo {year}
  {2009})}\BibitemShut {NoStop}%
\bibitem [{\citenamefont {Qi}\ and\ \citenamefont
  {Zhang}(2011)}]{r11_Qi_TIandSuperCon}%
  \BibitemOpen
  \bibfield  {author} {\bibinfo {author} {\bibfnamefont {X.-L.}\ \bibnamefont
  {Qi}}\ and\ \bibinfo {author} {\bibfnamefont {S.-C.}\ \bibnamefont {Zhang}},\
  }\bibfield  {title} {\bibinfo {title} {Topological insulators and
  superconductors},\ }\href {https://doi.org/10.1103/RevModPhys.83.1057}
  {\bibfield  {journal} {\bibinfo  {journal} {Rev. Mod. Phys.}\ }\textbf
  {\bibinfo {volume} {83}},\ \bibinfo {pages} {1057} (\bibinfo {year}
  {2011})}\BibitemShut {NoStop}%
\bibitem [{\citenamefont {Hsieh}\ \emph {et~al.}(2009)\citenamefont {Hsieh},
  \citenamefont {Xia}, \citenamefont {Qian}, \citenamefont {Wray},
  \citenamefont {Meier}, \citenamefont {Dil}, \citenamefont {Osterwalder},
  \citenamefont {Patthey}, \citenamefont {Fedorov}, \citenamefont {Lin},
  \citenamefont {Bansil}, \citenamefont {Grauer}, \citenamefont {Hor},
  \citenamefont {Cava},\ and\ \citenamefont {Hasan}}]{r09_Hsieh_ExpDiracCones}%
  \BibitemOpen
  \bibfield  {author} {\bibinfo {author} {\bibfnamefont {D.}~\bibnamefont
  {Hsieh}}, \bibinfo {author} {\bibfnamefont {Y.}~\bibnamefont {Xia}}, \bibinfo
  {author} {\bibfnamefont {D.}~\bibnamefont {Qian}}, \bibinfo {author}
  {\bibfnamefont {L.}~\bibnamefont {Wray}}, \bibinfo {author} {\bibfnamefont
  {F.}~\bibnamefont {Meier}}, \bibinfo {author} {\bibfnamefont {J.~H.}\
  \bibnamefont {Dil}}, \bibinfo {author} {\bibfnamefont {J.}~\bibnamefont
  {Osterwalder}}, \bibinfo {author} {\bibfnamefont {L.}~\bibnamefont
  {Patthey}}, \bibinfo {author} {\bibfnamefont {A.~V.}\ \bibnamefont
  {Fedorov}}, \bibinfo {author} {\bibfnamefont {H.}~\bibnamefont {Lin}},
  \bibinfo {author} {\bibfnamefont {A.}~\bibnamefont {Bansil}}, \bibinfo
  {author} {\bibfnamefont {D.}~\bibnamefont {Grauer}}, \bibinfo {author}
  {\bibfnamefont {Y.~S.}\ \bibnamefont {Hor}}, \bibinfo {author} {\bibfnamefont
  {R.~J.}\ \bibnamefont {Cava}},\ and\ \bibinfo {author} {\bibfnamefont
  {M.~Z.}\ \bibnamefont {Hasan}},\ }\bibfield  {title} {\bibinfo {title}
  {Observation of time-reversal-protected single-dirac-cone
  topological-insulator states in {${\mathrm{Bi}_{2}}{\mathrm{Te}}_{3}$} and
  {${\mathrm{Sb}}_{2}{\mathrm{Te}}_{3}$}},\ }\href
  {https://doi.org/10.1103/PhysRevLett.103.146401} {\bibfield  {journal}
  {\bibinfo  {journal} {Phys. Rev. Lett.}\ }\textbf {\bibinfo {volume} {103}},\
  \bibinfo {pages} {146401} (\bibinfo {year} {2009})}\BibitemShut {NoStop}%
\bibitem [{\citenamefont {Hasan}\ and\ \citenamefont
  {Kane}(2010)}]{r10_Hasan_TIreview}%
  \BibitemOpen
  \bibfield  {author} {\bibinfo {author} {\bibfnamefont {M.~Z.}\ \bibnamefont
  {Hasan}}\ and\ \bibinfo {author} {\bibfnamefont {C.~L.}\ \bibnamefont
  {Kane}},\ }\bibfield  {title} {\bibinfo {title} {Colloquium: Topological
  insulators},\ }\href {https://doi.org/10.1103/RevModPhys.82.3045} {\bibfield
  {journal} {\bibinfo  {journal} {Rev. Mod. Phys.}\ }\textbf {\bibinfo {volume}
  {82}},\ \bibinfo {pages} {3045} (\bibinfo {year} {2010})}\BibitemShut
  {NoStop}%
\bibitem [{\citenamefont {Zhang}\ \emph
  {et~al.}(2010{\natexlab{a}})\citenamefont {Zhang}, \citenamefont {He},
  \citenamefont {Chang}, \citenamefont {Song}, \citenamefont {Wang},
  \citenamefont {Chen}, \citenamefont {Jia}, \citenamefont {Fang},
  \citenamefont {Dai}, \citenamefont {Shan}, \citenamefont {Shen},
  \citenamefont {Niu}, \citenamefont {Qi}, \citenamefont {Zhang}, \citenamefont
  {Ma},\ and\ \citenamefont {Xue}}]{r10_Zhang_QLdependence}%
  \BibitemOpen
  \bibfield  {author} {\bibinfo {author} {\bibfnamefont {Y.}~\bibnamefont
  {Zhang}}, \bibinfo {author} {\bibfnamefont {K.}~\bibnamefont {He}}, \bibinfo
  {author} {\bibfnamefont {C.-Z.}\ \bibnamefont {Chang}}, \bibinfo {author}
  {\bibfnamefont {C.-L.}\ \bibnamefont {Song}}, \bibinfo {author}
  {\bibfnamefont {L.-L.}\ \bibnamefont {Wang}}, \bibinfo {author}
  {\bibfnamefont {X.~.}\ \bibnamefont {Chen}}, \bibinfo {author} {\bibfnamefont
  {J.-F.}\ \bibnamefont {Jia}}, \bibinfo {author} {\bibfnamefont
  {Z.}~\bibnamefont {Fang}}, \bibinfo {author} {\bibfnamefont {X.}~\bibnamefont
  {Dai}}, \bibinfo {author} {\bibfnamefont {W.-Y.}\ \bibnamefont {Shan}},
  \bibinfo {author} {\bibfnamefont {S.-Q.}\ \bibnamefont {Shen}}, \bibinfo
  {author} {\bibfnamefont {Q.}~\bibnamefont {Niu}}, \bibinfo {author}
  {\bibfnamefont {X.-L.}\ \bibnamefont {Qi}}, \bibinfo {author} {\bibfnamefont
  {S.-C.}\ \bibnamefont {Zhang}}, \bibinfo {author} {\bibfnamefont {X.-C.}\
  \bibnamefont {Ma}},\ and\ \bibinfo {author} {\bibfnamefont {Q.-K.}\
  \bibnamefont {Xue}},\ }\bibfield  {title} {\bibinfo {title} {Crossover of the
  three-dimensional topological insulator {Bi}$_2${Se}$_3$ to the
  two-dimensional limit},\ }\href {https://doi.org/10.1038/nphys1689}
  {\bibfield  {journal} {\bibinfo  {journal} {Nature Physics}\ }\textbf
  {\bibinfo {volume} {6}},\ \bibinfo {pages} {584 EP } (\bibinfo {year}
  {2010}{\natexlab{a}})}\BibitemShut {NoStop}%
\bibitem [{\citenamefont {Cayssol}(2013)}]{r13_Cayssol_DiracEq}%
  \BibitemOpen
  \bibfield  {author} {\bibinfo {author} {\bibfnamefont {J.}~\bibnamefont
  {Cayssol}},\ }\bibfield  {title} {\bibinfo {title} {Introduction to dirac
  materials and topological insulators},\ }\href
  {https://doi.org/10.1016/j.crhy.2013.09.012} {\bibfield  {journal} {\bibinfo
  {journal} {Comptes Rendus Physique}\ }\textbf {\bibinfo {volume} {14}},\
  \bibinfo {pages} {760 } (\bibinfo {year} {2013})},\ \bibinfo {note}
  {topological insulators / Isolants topologiques}\BibitemShut {NoStop}%
\bibitem [{\citenamefont {Bernevig}\ and\ \citenamefont
  {Hughes}(2013)}]{book_Bernevig_TI}%
  \BibitemOpen
  \bibfield  {author} {\bibinfo {author} {\bibfnamefont {B.}~\bibnamefont
  {Bernevig}}\ and\ \bibinfo {author} {\bibfnamefont {T.}~\bibnamefont
  {Hughes}},\ }\href {https://books.google.cz/books?id=wOn7JHSSxrsC} {\emph
  {\bibinfo {title} {Topological Insulators and Topological Superconductors}}}\
  (\bibinfo  {publisher} {Princeton University Press, New Jersey, USA},\
  \bibinfo {year} {2013})\BibitemShut {NoStop}%
\bibitem [{\citenamefont {Ortmann}\ \emph {et~al.}(2015)\citenamefont
  {Ortmann}, \citenamefont {Roche},\ and\ \citenamefont
  {Valenzuela}}]{r15_Ortmann_TI}%
  \BibitemOpen
  \bibfield  {author} {\bibinfo {author} {\bibfnamefont {F.}~\bibnamefont
  {Ortmann}}, \bibinfo {author} {\bibfnamefont {S.}~\bibnamefont {Roche}},\
  and\ \bibinfo {author} {\bibfnamefont {S.}~\bibnamefont {Valenzuela}},\
  }\href {https://books.google.cz/books?id=oph9rgEACAAJ} {\emph {\bibinfo
  {title} {Topological Insulators: Fundamentals and Perspectives}}}\ (\bibinfo
  {publisher} {Wiley-VCH, Weinheim, Germany},\ \bibinfo {year}
  {2015})\BibitemShut {NoStop}%
\bibitem [{\citenamefont {K{\"o}nig}\ \emph {et~al.}(2007)\citenamefont
  {K{\"o}nig}, \citenamefont {Wiedmann}, \citenamefont {Br{\"u}ne},
  \citenamefont {Roth}, \citenamefont {Buhmann}, \citenamefont {Molenkamp},
  \citenamefont {Qi},\ and\ \citenamefont {Zhang}}]{r07_Konig_QSHE}%
  \BibitemOpen
  \bibfield  {author} {\bibinfo {author} {\bibfnamefont {M.}~\bibnamefont
  {K{\"o}nig}}, \bibinfo {author} {\bibfnamefont {S.}~\bibnamefont {Wiedmann}},
  \bibinfo {author} {\bibfnamefont {C.}~\bibnamefont {Br{\"u}ne}}, \bibinfo
  {author} {\bibfnamefont {A.}~\bibnamefont {Roth}}, \bibinfo {author}
  {\bibfnamefont {H.}~\bibnamefont {Buhmann}}, \bibinfo {author} {\bibfnamefont
  {L.~W.}\ \bibnamefont {Molenkamp}}, \bibinfo {author} {\bibfnamefont {X.-L.}\
  \bibnamefont {Qi}},\ and\ \bibinfo {author} {\bibfnamefont {S.-C.}\
  \bibnamefont {Zhang}},\ }\bibfield  {title} {\bibinfo {title} {Quantum spin
  hall insulator state in hgte quantum wells},\ }\href
  {https://doi.org/10.1126/science.1148047} {\bibfield  {journal} {\bibinfo
  {journal} {Science}\ }\textbf {\bibinfo {volume} {318}},\ \bibinfo {pages}
  {766} (\bibinfo {year} {2007})},\ \Eprint
  {https://arxiv.org/abs/https://science.sciencemag.org/content/318/5851/-766.full.pdf}
  {https://science.sciencemag.org/content/318/5851/-766.full.pdf} \BibitemShut
  {NoStop}%
\bibitem [{\citenamefont {Wray}\ \emph {et~al.}(2010)\citenamefont {Wray},
  \citenamefont {Xu}, \citenamefont {Xia}, \citenamefont {Hsieh}, \citenamefont
  {Fedorov}, \citenamefont {Hor}, \citenamefont {Cava}, \citenamefont {Bansil},
  \citenamefont {Lin},\ and\ \citenamefont {Hasan}}]{r10_Wray_TI_magperturb}%
  \BibitemOpen
  \bibfield  {author} {\bibinfo {author} {\bibfnamefont {L.~A.}\ \bibnamefont
  {Wray}}, \bibinfo {author} {\bibfnamefont {S.-Y.}\ \bibnamefont {Xu}},
  \bibinfo {author} {\bibfnamefont {Y.}~\bibnamefont {Xia}}, \bibinfo {author}
  {\bibfnamefont {D.}~\bibnamefont {Hsieh}}, \bibinfo {author} {\bibfnamefont
  {A.~V.}\ \bibnamefont {Fedorov}}, \bibinfo {author} {\bibfnamefont {Y.~S.}\
  \bibnamefont {Hor}}, \bibinfo {author} {\bibfnamefont {R.~J.}\ \bibnamefont
  {Cava}}, \bibinfo {author} {\bibfnamefont {A.}~\bibnamefont {Bansil}},
  \bibinfo {author} {\bibfnamefont {H.}~\bibnamefont {Lin}},\ and\ \bibinfo
  {author} {\bibfnamefont {M.~Z.}\ \bibnamefont {Hasan}},\ }\bibfield  {title}
  {\bibinfo {title} {A topological insulator surface under strong coulomb,
  magnetic and disorder perturbations},\ }\href
  {https://doi.org/10.1038/nphys1838} {\bibfield  {journal} {\bibinfo
  {journal} {Nature Physics}\ }\textbf {\bibinfo {volume} {7}},\ \bibinfo
  {pages} {32 EP } (\bibinfo {year} {2010})}\BibitemShut {NoStop}%
\bibitem [{\citenamefont {Xu}\ \emph {et~al.}(2012)\citenamefont {Xu},
  \citenamefont {Neupane}, \citenamefont {Liu}, \citenamefont {Zhang},
  \citenamefont {Richardella}, \citenamefont {Andrew~Wray}, \citenamefont
  {Alidoust}, \citenamefont {Leandersson}, \citenamefont {Balasubramanian},
  \citenamefont {S{\'a}nchez-Barriga}, \citenamefont {Rader}, \citenamefont
  {Landolt}, \citenamefont {Slomski}, \citenamefont {Hugo~Dil}, \citenamefont
  {Osterwalder}, \citenamefont {Chang}, \citenamefont {Jeng}, \citenamefont
  {Lin}, \citenamefont {Bansil}, \citenamefont {Samarth},\ and\ \citenamefont
  {Zahid~Hasan}}]{r12_Xu_TImagdop}%
  \BibitemOpen
  \bibfield  {author} {\bibinfo {author} {\bibfnamefont {S.-Y.}\ \bibnamefont
  {Xu}}, \bibinfo {author} {\bibfnamefont {M.}~\bibnamefont {Neupane}},
  \bibinfo {author} {\bibfnamefont {C.}~\bibnamefont {Liu}}, \bibinfo {author}
  {\bibfnamefont {D.}~\bibnamefont {Zhang}}, \bibinfo {author} {\bibfnamefont
  {A.}~\bibnamefont {Richardella}}, \bibinfo {author} {\bibfnamefont
  {L.}~\bibnamefont {Andrew~Wray}}, \bibinfo {author} {\bibfnamefont
  {N.}~\bibnamefont {Alidoust}}, \bibinfo {author} {\bibfnamefont
  {M.}~\bibnamefont {Leandersson}}, \bibinfo {author} {\bibfnamefont
  {T.}~\bibnamefont {Balasubramanian}}, \bibinfo {author} {\bibfnamefont
  {J.}~\bibnamefont {S{\'a}nchez-Barriga}}, \bibinfo {author} {\bibfnamefont
  {O.}~\bibnamefont {Rader}}, \bibinfo {author} {\bibfnamefont
  {G.}~\bibnamefont {Landolt}}, \bibinfo {author} {\bibfnamefont
  {B.}~\bibnamefont {Slomski}}, \bibinfo {author} {\bibfnamefont
  {J.}~\bibnamefont {Hugo~Dil}}, \bibinfo {author} {\bibfnamefont
  {J.}~\bibnamefont {Osterwalder}}, \bibinfo {author} {\bibfnamefont {T.-R.}\
  \bibnamefont {Chang}}, \bibinfo {author} {\bibfnamefont {H.-T.}\ \bibnamefont
  {Jeng}}, \bibinfo {author} {\bibfnamefont {H.}~\bibnamefont {Lin}}, \bibinfo
  {author} {\bibfnamefont {A.}~\bibnamefont {Bansil}}, \bibinfo {author}
  {\bibfnamefont {N.}~\bibnamefont {Samarth}},\ and\ \bibinfo {author}
  {\bibfnamefont {M.}~\bibnamefont {Zahid~Hasan}},\ }\bibfield  {title}
  {\bibinfo {title} {Hedgehog spin texture and berry's phase tuning in a
  magnetic topological insulator},\ }\href {https://doi.org/10.1038/nphys2351}
  {\bibfield  {journal} {\bibinfo  {journal} {Nature Physics}\ }\textbf
  {\bibinfo {volume} {8}},\ \bibinfo {pages} {616 EP } (\bibinfo {year}
  {2012})},\ \bibinfo {note} {article}\BibitemShut {NoStop}%
\bibitem [{\citenamefont {Chang}\ \emph {et~al.}(2013)\citenamefont {Chang},
  \citenamefont {Zhang}, \citenamefont {Liu}, \citenamefont {Zhang},
  \citenamefont {Feng}, \citenamefont {Li}, \citenamefont {Wang}, \citenamefont
  {Chen}, \citenamefont {Dai}, \citenamefont {Fang}, \citenamefont {Qi},
  \citenamefont {Zhang}, \citenamefont {Wang}, \citenamefont {He},
  \citenamefont {Ma},\ and\ \citenamefont {Xue}}]{r13_Chang_QAHE}%
  \BibitemOpen
  \bibfield  {author} {\bibinfo {author} {\bibfnamefont {C.-Z.}\ \bibnamefont
  {Chang}}, \bibinfo {author} {\bibfnamefont {J.}~\bibnamefont {Zhang}},
  \bibinfo {author} {\bibfnamefont {M.}~\bibnamefont {Liu}}, \bibinfo {author}
  {\bibfnamefont {Z.}~\bibnamefont {Zhang}}, \bibinfo {author} {\bibfnamefont
  {X.}~\bibnamefont {Feng}}, \bibinfo {author} {\bibfnamefont {K.}~\bibnamefont
  {Li}}, \bibinfo {author} {\bibfnamefont {L.-L.}\ \bibnamefont {Wang}},
  \bibinfo {author} {\bibfnamefont {X.}~\bibnamefont {Chen}}, \bibinfo {author}
  {\bibfnamefont {X.}~\bibnamefont {Dai}}, \bibinfo {author} {\bibfnamefont
  {Z.}~\bibnamefont {Fang}}, \bibinfo {author} {\bibfnamefont {X.-L.}\
  \bibnamefont {Qi}}, \bibinfo {author} {\bibfnamefont {S.-C.}\ \bibnamefont
  {Zhang}}, \bibinfo {author} {\bibfnamefont {Y.}~\bibnamefont {Wang}},
  \bibinfo {author} {\bibfnamefont {K.}~\bibnamefont {He}}, \bibinfo {author}
  {\bibfnamefont {X.-C.}\ \bibnamefont {Ma}},\ and\ \bibinfo {author}
  {\bibfnamefont {Q.-K.}\ \bibnamefont {Xue}},\ }\bibfield  {title} {\bibinfo
  {title} {Thin films of magnetically doped topological insulator with
  carrier-independent long-range ferromagnetic order},\ }\href
  {https://doi.org/10.1002/adma.201203493} {\bibfield  {journal} {\bibinfo
  {journal} {Advanced Materials}\ }\textbf {\bibinfo {volume} {25}},\ \bibinfo
  {pages} {1065} (\bibinfo {year} {2013})},\ \Eprint
  {https://arxiv.org/abs/https://onlinelibrary.wiley.com/doi/pdf/10.1002/-adma.201203493}
  {https://onlinelibrary.wiley.com/doi/pdf/10.1002/-adma.201203493}
  \BibitemShut {NoStop}%
\bibitem [{\citenamefont {Carva}\ \emph {et~al.}(2016)\citenamefont {Carva},
  \citenamefont {Kudrnovsk\'y}, \citenamefont {M\'aca}, \citenamefont {Drchal},
  \citenamefont {Turek}, \citenamefont {Bal\'a\ifmmode~\check{z}\else
  \v{z}\fi{}}, \citenamefont {Tk\'a\ifmmode~\check{c}\else \v{c}\fi{}},
  \citenamefont {Hol\'y}, \citenamefont {Sechovsk\'y},\ and\ \citenamefont
  {Honolka}}]{r16_Carva_BiTe}%
  \BibitemOpen
  \bibfield  {author} {\bibinfo {author} {\bibfnamefont {K.}~\bibnamefont
  {Carva}}, \bibinfo {author} {\bibfnamefont {J.}~\bibnamefont {Kudrnovsk\'y}},
  \bibinfo {author} {\bibfnamefont {F.}~\bibnamefont {M\'aca}}, \bibinfo
  {author} {\bibfnamefont {V.}~\bibnamefont {Drchal}}, \bibinfo {author}
  {\bibfnamefont {I.}~\bibnamefont {Turek}}, \bibinfo {author} {\bibfnamefont
  {P.}~\bibnamefont {Bal\'a\ifmmode~\check{z}\else \v{z}\fi{}}}, \bibinfo
  {author} {\bibfnamefont {V.}~\bibnamefont {Tk\'a\ifmmode~\check{c}\else
  \v{c}\fi{}}}, \bibinfo {author} {\bibfnamefont {V.}~\bibnamefont {Hol\'y}},
  \bibinfo {author} {\bibfnamefont {V.}~\bibnamefont {Sechovsk\'y}},\ and\
  \bibinfo {author} {\bibfnamefont {J.}~\bibnamefont {Honolka}},\ }\bibfield
  {title} {\bibinfo {title} {Electronic and transport properties of the
  {Mn}-doped topological insulator {${\mathrm{Bi}}_{2}{\mathrm{Te}}_{3}$}: A
  first-principles study},\ }\href {https://doi.org/10.1103/PhysRevB.93.214409}
  {\bibfield  {journal} {\bibinfo  {journal} {Phys. Rev. B}\ }\textbf {\bibinfo
  {volume} {93}},\ \bibinfo {pages} {214409} (\bibinfo {year}
  {2016})}\BibitemShut {NoStop}%
\bibitem [{\citenamefont {Máca}\ \emph {et~al.}(2019)\citenamefont {Máca},
  \citenamefont {Kudrnovský}, \citenamefont {Baláž}, \citenamefont {Drchal},
  \citenamefont {Carva},\ and\ \citenamefont {Turek}}]{r19_Maca_CuMnAs}%
  \BibitemOpen
  \bibfield  {author} {\bibinfo {author} {\bibfnamefont {F.}~\bibnamefont
  {Máca}}, \bibinfo {author} {\bibfnamefont {J.}~\bibnamefont {Kudrnovský}},
  \bibinfo {author} {\bibfnamefont {P.}~\bibnamefont {Baláž}}, \bibinfo
  {author} {\bibfnamefont {V.}~\bibnamefont {Drchal}}, \bibinfo {author}
  {\bibfnamefont {K.}~\bibnamefont {Carva}},\ and\ \bibinfo {author}
  {\bibfnamefont {I.}~\bibnamefont {Turek}},\ }\bibfield  {title} {\bibinfo
  {title} {{Tetragonal {C}u{Mn}{As} alloy: Role of defects}},\ }\href
  {https://doi.org/10.1016/j.jmmm.2018.10.145} {\bibfield  {journal} {\bibinfo
  {journal} {Journal of Magnetism and Magnetic Materials}\ }\textbf {\bibinfo
  {volume} {474}},\ \bibinfo {pages} {467 } (\bibinfo {year}
  {2019})}\BibitemShut {NoStop}%
\bibitem [{\citenamefont {Zhang}\ \emph {et~al.}(2009)\citenamefont {Zhang},
  \citenamefont {Liu}, \citenamefont {Qi}, \citenamefont {Dai}, \citenamefont
  {Fang},\ and\ \citenamefont {Zhang}}]{r09_Zhang_Bi2x3}%
  \BibitemOpen
  \bibfield  {author} {\bibinfo {author} {\bibfnamefont {H.}~\bibnamefont
  {Zhang}}, \bibinfo {author} {\bibfnamefont {C.-X.}\ \bibnamefont {Liu}},
  \bibinfo {author} {\bibfnamefont {X.-L.}\ \bibnamefont {Qi}}, \bibinfo
  {author} {\bibfnamefont {X.}~\bibnamefont {Dai}}, \bibinfo {author}
  {\bibfnamefont {Z.}~\bibnamefont {Fang}},\ and\ \bibinfo {author}
  {\bibfnamefont {S.-C.}\ \bibnamefont {Zhang}},\ }\bibfield  {title} {\bibinfo
  {title} {Topological insulators in {B}i$_{2}${S}e$_{3}$, {B}i$_{2}${T}e$_{3}$
  and {S}b$_{2}${T}e$_{3}$ with a single dirac cone on the surface},\ }\href
  {https://doi.org/10.1038/nphys1270} {\bibfield  {journal} {\bibinfo
  {journal} {Nature Physics}\ }\textbf {\bibinfo {volume} {5}},\ \bibinfo
  {pages} {438 EP } (\bibinfo {year} {2009})},\ \bibinfo {note}
  {article}\BibitemShut {NoStop}%
\bibitem [{\citenamefont {Zhang}\ \emph
  {et~al.}(2010{\natexlab{b}})\citenamefont {Zhang}, \citenamefont {Yu},
  \citenamefont {Zhang}, \citenamefont {Dai},\ and\ \citenamefont
  {Fang}}]{r10_Zhang_2010}%
  \BibitemOpen
  \bibfield  {author} {\bibinfo {author} {\bibfnamefont {W.}~\bibnamefont
  {Zhang}}, \bibinfo {author} {\bibfnamefont {R.}~\bibnamefont {Yu}}, \bibinfo
  {author} {\bibfnamefont {H.-J.}\ \bibnamefont {Zhang}}, \bibinfo {author}
  {\bibfnamefont {X.}~\bibnamefont {Dai}},\ and\ \bibinfo {author}
  {\bibfnamefont {Z.}~\bibnamefont {Fang}},\ }\bibfield  {title} {\bibinfo
  {title} {First-principles studies of the three-dimensional strong topological
  insulators {B}i$_{2}${T}e$_{3}$, {B}i$_{2}${S}e$_{3}$ and
  {S}b$_{2}${T}e$_{3}$},\ }\href
  {https://doi.org/10.1088/1367-2630/12/6/065013} {\bibfield  {journal}
  {\bibinfo  {journal} {New Journal of Physics}\ }\textbf {\bibinfo {volume}
  {12}},\ \bibinfo {pages} {065013} (\bibinfo {year}
  {2010}{\natexlab{b}})}\BibitemShut {NoStop}%
\bibitem [{\citenamefont {Lee}\ \emph {et~al.}(2014)\citenamefont {Lee},
  \citenamefont {Punugupati}, \citenamefont {Wu}, \citenamefont {Jin},
  \citenamefont {Narayan},\ and\ \citenamefont
  {Schwartz}}]{r14_Lee_BiSe_TP_SeDeficit}%
  \BibitemOpen
  \bibfield  {author} {\bibinfo {author} {\bibfnamefont {Y.}~\bibnamefont
  {Lee}}, \bibinfo {author} {\bibfnamefont {S.}~\bibnamefont {Punugupati}},
  \bibinfo {author} {\bibfnamefont {F.}~\bibnamefont {Wu}}, \bibinfo {author}
  {\bibfnamefont {Z.}~\bibnamefont {Jin}}, \bibinfo {author} {\bibfnamefont
  {J.}~\bibnamefont {Narayan}},\ and\ \bibinfo {author} {\bibfnamefont
  {J.}~\bibnamefont {Schwartz}},\ }\bibfield  {title} {\bibinfo {title}
  {Evidence for topological surface states in epitaxial {Bi}$_2${Se}$_3$ thin
  film grown by pulsed laser deposition through magneto-transport
  measurements},\ }\href {https://doi.org/10.1016/j.cossms.2014.07.001}
  {\bibfield  {journal} {\bibinfo  {journal} {Current Opinion in Solid State
  and Materials Science}\ }\textbf {\bibinfo {volume} {18}},\ \bibinfo {pages}
  {279 } (\bibinfo {year} {2014})}\BibitemShut {NoStop}%
\bibitem [{\citenamefont {Kriegner}\ \emph {et~al.}(2017)\citenamefont
  {Kriegner}, \citenamefont {Harcuba}, \citenamefont {Vesel{\'{y}}},
  \citenamefont {Lesnik}, \citenamefont {Bauer}, \citenamefont {Springholz},\
  and\ \citenamefont {Hol{\'{y}}}}]{r19_Kriegner_TP}%
  \BibitemOpen
  \bibfield  {author} {\bibinfo {author} {\bibfnamefont {D.}~\bibnamefont
  {Kriegner}}, \bibinfo {author} {\bibfnamefont {P.}~\bibnamefont {Harcuba}},
  \bibinfo {author} {\bibfnamefont {J.}~\bibnamefont {Vesel{\'{y}}}}, \bibinfo
  {author} {\bibfnamefont {A.}~\bibnamefont {Lesnik}}, \bibinfo {author}
  {\bibfnamefont {G.}~\bibnamefont {Bauer}}, \bibinfo {author} {\bibfnamefont
  {G.}~\bibnamefont {Springholz}},\ and\ \bibinfo {author} {\bibfnamefont
  {V.}~\bibnamefont {Hol{\'{y}}}},\ }\bibfield  {title} {\bibinfo {title}
  {{Twin domain imaging in topological insulator Bi${\sb 2}$Te${\sb 3}$ and
  Bi${\sb 2}$Se${\sb 3}$ epitaxial thin films by scanning X-ray nanobeam
  microscopy and electron backscatter diffraction}},\ }\href
  {https://doi.org/10.1107/S1600576717000565} {\bibfield  {journal} {\bibinfo
  {journal} {Journal of Applied Crystallography}\ }\textbf {\bibinfo {volume}
  {50}},\ \bibinfo {pages} {369} (\bibinfo {year} {2017})}\BibitemShut
  {NoStop}%
\bibitem [{\citenamefont {Zhang}\ \emph {et~al.}(2013)\citenamefont {Zhang},
  \citenamefont {Ming}, \citenamefont {Huang}, \citenamefont {Liu},
  \citenamefont {Kou}, \citenamefont {Fan}, \citenamefont {Wang},\ and\
  \citenamefont {Yao}}]{r13_Zhang_DopantsPos}%
  \BibitemOpen
  \bibfield  {author} {\bibinfo {author} {\bibfnamefont {J.-M.}\ \bibnamefont
  {Zhang}}, \bibinfo {author} {\bibfnamefont {W.}~\bibnamefont {Ming}},
  \bibinfo {author} {\bibfnamefont {Z.}~\bibnamefont {Huang}}, \bibinfo
  {author} {\bibfnamefont {G.-B.}\ \bibnamefont {Liu}}, \bibinfo {author}
  {\bibfnamefont {X.}~\bibnamefont {Kou}}, \bibinfo {author} {\bibfnamefont
  {Y.}~\bibnamefont {Fan}}, \bibinfo {author} {\bibfnamefont {K.~L.}\
  \bibnamefont {Wang}},\ and\ \bibinfo {author} {\bibfnamefont
  {Y.}~\bibnamefont {Yao}},\ }\bibfield  {title} {\bibinfo {title} {Stability,
  electronic, and magnetic properties of the magnetically doped topological
  insulators {Bi}$_{2}${Se}$_{3}$, {Bi}$_{2}${Te}$_{3}$, and
  {Sb}$_{2}${Te}$_{3}$},\ }\href {https://doi.org/10.1103/PhysRevB.88.235131}
  {\bibfield  {journal} {\bibinfo  {journal} {Phys. Rev. B}\ }\textbf {\bibinfo
  {volume} {88}},\ \bibinfo {pages} {235131} (\bibinfo {year}
  {2013})}\BibitemShut {NoStop}%
\bibitem [{\citenamefont {Hor}\ \emph {et~al.}(2010)\citenamefont {Hor},
  \citenamefont {Roushan}, \citenamefont {Beidenkopf}, \citenamefont {Seo},
  \citenamefont {Qu}, \citenamefont {Checkelsky}, \citenamefont {Wray},
  \citenamefont {Hsieh}, \citenamefont {Xia}, \citenamefont {Xu}, \citenamefont
  {Qian}, \citenamefont {Hasan}, \citenamefont {Ong}, \citenamefont {Yazdani},\
  and\ \citenamefont {Cava}}]{r10_Hor_DopantsPosMn}%
  \BibitemOpen
  \bibfield  {author} {\bibinfo {author} {\bibfnamefont {Y.~S.}\ \bibnamefont
  {Hor}}, \bibinfo {author} {\bibfnamefont {P.}~\bibnamefont {Roushan}},
  \bibinfo {author} {\bibfnamefont {H.}~\bibnamefont {Beidenkopf}}, \bibinfo
  {author} {\bibfnamefont {J.}~\bibnamefont {Seo}}, \bibinfo {author}
  {\bibfnamefont {D.}~\bibnamefont {Qu}}, \bibinfo {author} {\bibfnamefont
  {J.~G.}\ \bibnamefont {Checkelsky}}, \bibinfo {author} {\bibfnamefont
  {L.~A.}\ \bibnamefont {Wray}}, \bibinfo {author} {\bibfnamefont
  {D.}~\bibnamefont {Hsieh}}, \bibinfo {author} {\bibfnamefont
  {Y.}~\bibnamefont {Xia}}, \bibinfo {author} {\bibfnamefont {S.-Y.}\
  \bibnamefont {Xu}}, \bibinfo {author} {\bibfnamefont {D.}~\bibnamefont
  {Qian}}, \bibinfo {author} {\bibfnamefont {M.~Z.}\ \bibnamefont {Hasan}},
  \bibinfo {author} {\bibfnamefont {N.~P.}\ \bibnamefont {Ong}}, \bibinfo
  {author} {\bibfnamefont {A.}~\bibnamefont {Yazdani}},\ and\ \bibinfo {author}
  {\bibfnamefont {R.~J.}\ \bibnamefont {Cava}},\ }\bibfield  {title} {\bibinfo
  {title} {Development of ferromagnetism in the doped topological insulator
  {Bi}$_{\mathrm{2\ensuremath{-}x}}${Mn}$_\mathrm{{x}}${{Te}}$_{3}$},\ }\href
  {https://doi.org/10.1103/PhysRevB.81.195203} {\bibfield  {journal} {\bibinfo
  {journal} {Phys. Rev. B}\ }\textbf {\bibinfo {volume} {81}},\ \bibinfo
  {pages} {195203} (\bibinfo {year} {2010})}\BibitemShut {NoStop}%
\bibitem [{\citenamefont {Zhang}\ \emph {et~al.}(2012)\citenamefont {Zhang},
  \citenamefont {Zhu}, \citenamefont {Zhang}, \citenamefont {Xiao},\ and\
  \citenamefont {Yao}}]{r12_Zhang_dopants}%
  \BibitemOpen
  \bibfield  {author} {\bibinfo {author} {\bibfnamefont {J.-M.}\ \bibnamefont
  {Zhang}}, \bibinfo {author} {\bibfnamefont {W.}~\bibnamefont {Zhu}}, \bibinfo
  {author} {\bibfnamefont {Y.}~\bibnamefont {Zhang}}, \bibinfo {author}
  {\bibfnamefont {D.}~\bibnamefont {Xiao}},\ and\ \bibinfo {author}
  {\bibfnamefont {Y.}~\bibnamefont {Yao}},\ }\bibfield  {title} {\bibinfo
  {title} {Tailoring magnetic doping in the topological insulator
  {Bi}$_{2}${Se}$_{3}$},\ }\href
  {https://doi.org/10.1103/PhysRevLett.109.266405} {\bibfield  {journal}
  {\bibinfo  {journal} {Phys. Rev. Lett.}\ }\textbf {\bibinfo {volume} {109}},\
  \bibinfo {pages} {266405} (\bibinfo {year} {2012})}\BibitemShut {NoStop}%
\bibitem [{\citenamefont {Rienks}\ \emph {et~al.}(2019)\citenamefont {Rienks},
  \citenamefont {Wimmer}, \citenamefont {S{\'a}nchez-Barriga}, \citenamefont
  {Caha}, \citenamefont {Mandal}, \citenamefont {Ruzicka}, \citenamefont {Ney},
  \citenamefont {Steiner}, \citenamefont {Volobuev}, \citenamefont {Groiss},
  \citenamefont {Albu}, \citenamefont {Kothleitner}, \citenamefont {Khan},
  \citenamefont {Min{\'a}r}, \citenamefont {Ebert}, \citenamefont {Bauer},
  \citenamefont {Freyse}, \citenamefont {Varykhalov}, \citenamefont {Rader},\
  and\ \citenamefont {Springholz}}]{r18ArXiv_Rienks_Mnsextuplety}%
  \BibitemOpen
  \bibfield  {author} {\bibinfo {author} {\bibfnamefont {E.~D.~L.}\
  \bibnamefont {Rienks}}, \bibinfo {author} {\bibfnamefont {S.}~\bibnamefont
  {Wimmer}}, \bibinfo {author} {\bibfnamefont {J.}~\bibnamefont
  {S{\'a}nchez-Barriga}}, \bibinfo {author} {\bibfnamefont {O.}~\bibnamefont
  {Caha}}, \bibinfo {author} {\bibfnamefont {P.~S.}\ \bibnamefont {Mandal}},
  \bibinfo {author} {\bibfnamefont {J.}~\bibnamefont {Ruzicka}}, \bibinfo
  {author} {\bibfnamefont {A.}~\bibnamefont {Ney}}, \bibinfo {author}
  {\bibfnamefont {H.}~\bibnamefont {Steiner}}, \bibinfo {author} {\bibfnamefont
  {V.~V.}\ \bibnamefont {Volobuev}}, \bibinfo {author} {\bibfnamefont
  {H.}~\bibnamefont {Groiss}}, \bibinfo {author} {\bibfnamefont
  {M.}~\bibnamefont {Albu}}, \bibinfo {author} {\bibfnamefont {J.}~\bibnamefont
  {Kothleitner}, \bibfnamefont {G.and~Michalicka}}, \bibinfo {author}
  {\bibfnamefont {S.~A.}\ \bibnamefont {Khan}}, \bibinfo {author}
  {\bibfnamefont {J.}~\bibnamefont {Min{\'a}r}}, \bibinfo {author}
  {\bibfnamefont {H.}~\bibnamefont {Ebert}}, \bibinfo {author} {\bibfnamefont
  {G.}~\bibnamefont {Bauer}}, \bibinfo {author} {\bibfnamefont
  {F.}~\bibnamefont {Freyse}}, \bibinfo {author} {\bibfnamefont
  {A.}~\bibnamefont {Varykhalov}}, \bibinfo {author} {\bibfnamefont
  {O.}~\bibnamefont {Rader}},\ and\ \bibinfo {author} {\bibfnamefont
  {G.}~\bibnamefont {Springholz}},\ }\bibfield  {title} {\bibinfo {title}
  {Large magnetic gap at the dirac point in
  {Bi}$_2${Te}$_3$/{Mn}{Bi}$_2${Te}$_4$ heterostructures},\ }\href
  {https://doi.org/10.1038/s41586-019-1826-7} {\bibfield  {journal} {\bibinfo
  {journal} {Nature}\ }\textbf {\bibinfo {volume} {576}},\ \bibinfo {pages}
  {423} (\bibinfo {year} {2019})}\BibitemShut {NoStop}%
\bibitem [{\citenamefont {Hor}\ \emph {et~al.}(2009)\citenamefont {Hor},
  \citenamefont {Richardella}, \citenamefont {Roushan}, \citenamefont {Xia},
  \citenamefont {Checkelsky}, \citenamefont {Yazdani}, \citenamefont {Hasan},
  \citenamefont {Ong},\ and\ \citenamefont {Cava}}]{r09_Hor_p-type_Bi2Se3_Ca}%
  \BibitemOpen
  \bibfield  {author} {\bibinfo {author} {\bibfnamefont {Y.~S.}\ \bibnamefont
  {Hor}}, \bibinfo {author} {\bibfnamefont {A.}~\bibnamefont {Richardella}},
  \bibinfo {author} {\bibfnamefont {P.}~\bibnamefont {Roushan}}, \bibinfo
  {author} {\bibfnamefont {Y.}~\bibnamefont {Xia}}, \bibinfo {author}
  {\bibfnamefont {J.~G.}\ \bibnamefont {Checkelsky}}, \bibinfo {author}
  {\bibfnamefont {A.}~\bibnamefont {Yazdani}}, \bibinfo {author} {\bibfnamefont
  {M.~Z.}\ \bibnamefont {Hasan}}, \bibinfo {author} {\bibfnamefont {N.~P.}\
  \bibnamefont {Ong}},\ and\ \bibinfo {author} {\bibfnamefont {R.~J.}\
  \bibnamefont {Cava}},\ }\bibfield  {title} {\bibinfo {title} {$p$-type
  {Bi}$_{2}${Se}$_{3}$ for topological insulator and low-temperature
  thermoelectric applications},\ }\href
  {https://doi.org/10.1103/PhysRevB.79.195208} {\bibfield  {journal} {\bibinfo
  {journal} {Phys. Rev. B}\ }\textbf {\bibinfo {volume} {79}},\ \bibinfo
  {pages} {195208} (\bibinfo {year} {2009})}\BibitemShut {NoStop}%
\bibitem [{\citenamefont {Scanlon}\ \emph {et~al.}(2012)\citenamefont
  {Scanlon}, \citenamefont {King}, \citenamefont {Singh}, \citenamefont {de~la
  Torre}, \citenamefont {Walker}, \citenamefont {Balakrishnan}, \citenamefont
  {Baumberger},\ and\ \citenamefont {Catlow}}]{r12_Scanlon_TIantisites}%
  \BibitemOpen
  \bibfield  {author} {\bibinfo {author} {\bibfnamefont {D.~O.}\ \bibnamefont
  {Scanlon}}, \bibinfo {author} {\bibfnamefont {P.~D.~C.}\ \bibnamefont
  {King}}, \bibinfo {author} {\bibfnamefont {R.~P.}\ \bibnamefont {Singh}},
  \bibinfo {author} {\bibfnamefont {A.}~\bibnamefont {de~la Torre}}, \bibinfo
  {author} {\bibfnamefont {S.~M.}\ \bibnamefont {Walker}}, \bibinfo {author}
  {\bibfnamefont {G.}~\bibnamefont {Balakrishnan}}, \bibinfo {author}
  {\bibfnamefont {F.}~\bibnamefont {Baumberger}},\ and\ \bibinfo {author}
  {\bibfnamefont {C.~R.~A.}\ \bibnamefont {Catlow}},\ }\bibfield  {title}
  {\bibinfo {title} {Controlling bulk conductivity in topological insulators:
  Key role of anti-site defects},\ }\href
  {https://doi.org/10.1002/adma.201200187} {\bibfield  {journal} {\bibinfo
  {journal} {Advanced Materials}\ }\textbf {\bibinfo {volume} {24}},\ \bibinfo
  {pages} {2154} (\bibinfo {year} {2012})},\ \Eprint
  {https://arxiv.org/abs/https://onlinelibrary.wiley.com/doi/pdf/10.1002/-adma.201200187}
  {https://onlinelibrary.wiley.com/doi/pdf/10.1002/-adma.201200187}
  \BibitemShut {NoStop}%
\bibitem [{\citenamefont {Wolos}\ \emph {et~al.}(2016)\citenamefont {Wolos},
  \citenamefont {Drabinska}, \citenamefont {Borysiuk}, \citenamefont {Sobczak},
  \citenamefont {Kaminska}, \citenamefont {Hruban}, \citenamefont {Strzelecka},
  \citenamefont {Materna}, \citenamefont {Piersa}, \citenamefont {Romaniec},\
  and\ \citenamefont {Diduszko}}]{r16_Wolos_BiSevac}%
  \BibitemOpen
  \bibfield  {author} {\bibinfo {author} {\bibfnamefont {A.}~\bibnamefont
  {Wolos}}, \bibinfo {author} {\bibfnamefont {A.}~\bibnamefont {Drabinska}},
  \bibinfo {author} {\bibfnamefont {J.}~\bibnamefont {Borysiuk}}, \bibinfo
  {author} {\bibfnamefont {K.}~\bibnamefont {Sobczak}}, \bibinfo {author}
  {\bibfnamefont {M.}~\bibnamefont {Kaminska}}, \bibinfo {author}
  {\bibfnamefont {A.}~\bibnamefont {Hruban}}, \bibinfo {author} {\bibfnamefont
  {S.~G.}\ \bibnamefont {Strzelecka}}, \bibinfo {author} {\bibfnamefont
  {A.}~\bibnamefont {Materna}}, \bibinfo {author} {\bibfnamefont
  {M.}~\bibnamefont {Piersa}}, \bibinfo {author} {\bibfnamefont
  {M.}~\bibnamefont {Romaniec}},\ and\ \bibinfo {author} {\bibfnamefont
  {R.}~\bibnamefont {Diduszko}},\ }\bibfield  {title} {\bibinfo {title}
  {High-spin configuration of mn in {Bi}$_2${Se}$_3$ three-dimensional
  topological insulator},\ }\href {https://doi.org/10.1016/j.jmmm.2016.06.017}
  {\bibfield  {journal} {\bibinfo  {journal} {Journal of Magnetism and Magnetic
  Materials}\ }\textbf {\bibinfo {volume} {419}},\ \bibinfo {pages} {301 }
  (\bibinfo {year} {2016})}\BibitemShut {NoStop}%
\bibitem [{\citenamefont {Huang}\ \emph {et~al.}(2012)\citenamefont {Huang},
  \citenamefont {Chu}, \citenamefont {Kung}, \citenamefont {Lee}, \citenamefont
  {Sankar}, \citenamefont {Liou}, \citenamefont {Wu}, \citenamefont {Kuo},\
  and\ \citenamefont {Chou}}]{r18_Huang_Bi_antisites}%
  \BibitemOpen
  \bibfield  {author} {\bibinfo {author} {\bibfnamefont {F.-T.}\ \bibnamefont
  {Huang}}, \bibinfo {author} {\bibfnamefont {M.-W.}\ \bibnamefont {Chu}},
  \bibinfo {author} {\bibfnamefont {H.~H.}\ \bibnamefont {Kung}}, \bibinfo
  {author} {\bibfnamefont {W.~L.}\ \bibnamefont {Lee}}, \bibinfo {author}
  {\bibfnamefont {R.}~\bibnamefont {Sankar}}, \bibinfo {author} {\bibfnamefont
  {S.-C.}\ \bibnamefont {Liou}}, \bibinfo {author} {\bibfnamefont {K.~K.}\
  \bibnamefont {Wu}}, \bibinfo {author} {\bibfnamefont {Y.~K.}\ \bibnamefont
  {Kuo}},\ and\ \bibinfo {author} {\bibfnamefont {F.~C.}\ \bibnamefont
  {Chou}},\ }\bibfield  {title} {\bibinfo {title} {Nonstoichiometric doping and
  bi antisite defect in single crystal bi${}_{2}$se${}_{3}$},\ }\href
  {https://doi.org/10.1103/PhysRevB.86.081104} {\bibfield  {journal} {\bibinfo
  {journal} {Phys. Rev. B}\ }\textbf {\bibinfo {volume} {86}},\ \bibinfo
  {pages} {081104} (\bibinfo {year} {2012})}\BibitemShut {NoStop}%
\bibitem [{\citenamefont {Medlin}\ and\ \citenamefont
  {Yang}(2012)}]{r12_Medlin_TPaStep}%
  \BibitemOpen
  \bibfield  {author} {\bibinfo {author} {\bibfnamefont {D.~L.}\ \bibnamefont
  {Medlin}}\ and\ \bibinfo {author} {\bibfnamefont {N.~Y.~C.}\ \bibnamefont
  {Yang}},\ }\bibfield  {title} {\bibinfo {title} {{Interfacial Step Structure
  at a (0001) Basal Twin in {B}i$_{2}${T}e$_{3}$}},\ }\href
  {https://doi.org/10.1007/s11664-011-1859-7} {\bibfield  {journal} {\bibinfo
  {journal} {Journal of Electronic Materials}\ }\textbf {\bibinfo {volume}
  {41}},\ \bibinfo {pages} {1456} (\bibinfo {year} {2012})}\BibitemShut
  {NoStop}%
\bibitem [{\citenamefont {Medlin}\ \emph {et~al.}(2010)\citenamefont {Medlin},
  \citenamefont {Ramasse}, \citenamefont {Spataru},\ and\ \citenamefont
  {Yang}}]{r10_Medlin_TPenergie}%
  \BibitemOpen
  \bibfield  {author} {\bibinfo {author} {\bibfnamefont {D.~L.}\ \bibnamefont
  {Medlin}}, \bibinfo {author} {\bibfnamefont {Q.~M.}\ \bibnamefont {Ramasse}},
  \bibinfo {author} {\bibfnamefont {C.~D.}\ \bibnamefont {Spataru}},\ and\
  \bibinfo {author} {\bibfnamefont {N.~Y.~C.}\ \bibnamefont {Yang}},\
  }\bibfield  {title} {\bibinfo {title} {Structure of the (0001) basal twin
  boundary in {Bi$_2$Te$_3$}},\ }\href {https://doi.org/10.1063/1.3457902}
  {\bibfield  {journal} {\bibinfo  {journal} {Journal of Applied Physics}\
  }\textbf {\bibinfo {volume} {108}},\ \bibinfo {pages} {043517} (\bibinfo
  {year} {2010})},\ \Eprint
  {https://arxiv.org/abs/https://doi.org/10.1063/1.3457902}
  {https://doi.org/10.1063/1.3457902} \BibitemShut {NoStop}%
\bibitem [{\citenamefont {Tarakina}\ \emph {et~al.}(2014)\citenamefont
  {Tarakina}, \citenamefont {Schreyeck}, \citenamefont {Luysberg},
  \citenamefont {Grauer}, \citenamefont {Schumacher}, \citenamefont
  {Karczewski}, \citenamefont {Brunner}, \citenamefont {Gould}, \citenamefont
  {Buhmann}, \citenamefont {Dunin-Borkowski},\ and\ \citenamefont
  {Molenkamp}}]{r14_Tarakina_TPvsSubstrat}%
  \BibitemOpen
  \bibfield  {author} {\bibinfo {author} {\bibfnamefont {N.~V.}\ \bibnamefont
  {Tarakina}}, \bibinfo {author} {\bibfnamefont {S.}~\bibnamefont {Schreyeck}},
  \bibinfo {author} {\bibfnamefont {M.}~\bibnamefont {Luysberg}}, \bibinfo
  {author} {\bibfnamefont {S.}~\bibnamefont {Grauer}}, \bibinfo {author}
  {\bibfnamefont {C.}~\bibnamefont {Schumacher}}, \bibinfo {author}
  {\bibfnamefont {G.}~\bibnamefont {Karczewski}}, \bibinfo {author}
  {\bibfnamefont {K.}~\bibnamefont {Brunner}}, \bibinfo {author} {\bibfnamefont
  {C.}~\bibnamefont {Gould}}, \bibinfo {author} {\bibfnamefont
  {H.}~\bibnamefont {Buhmann}}, \bibinfo {author} {\bibfnamefont {R.~E.}\
  \bibnamefont {Dunin-Borkowski}},\ and\ \bibinfo {author} {\bibfnamefont
  {L.~W.}\ \bibnamefont {Molenkamp}},\ }\bibfield  {title} {\bibinfo {title}
  {Suppressing twin formation in {{Bi$_2$Se$_3$}} thin films},\ }\href
  {https://doi.org/10.1002/admi.201400134} {\bibfield  {journal} {\bibinfo
  {journal} {Advanced Materials Interfaces}\ }\textbf {\bibinfo {volume} {1}},\
  \bibinfo {pages} {1400134} (\bibinfo {year} {2014})},\ \Eprint
  {https://arxiv.org/abs/https://onlinelibrary.wiley.com/doi/pdf/10.1002/-admi.201400134}
  {https://onlinelibrary.wiley.com/doi/pdf/10.1002/-admi.201400134}
  \BibitemShut {NoStop}%
\bibitem [{\citenamefont {Levy}\ \emph {et~al.}(2018)\citenamefont {Levy},
  \citenamefont {Garcia}, \citenamefont {Shafique},\ and\ \citenamefont
  {Tamargo}}]{r18_Levy_PreGrowProcess}%
  \BibitemOpen
  \bibfield  {author} {\bibinfo {author} {\bibfnamefont {I.}~\bibnamefont
  {Levy}}, \bibinfo {author} {\bibfnamefont {T.~A.}\ \bibnamefont {Garcia}},
  \bibinfo {author} {\bibfnamefont {S.}~\bibnamefont {Shafique}},\ and\
  \bibinfo {author} {\bibfnamefont {M.~C.}\ \bibnamefont {Tamargo}},\
  }\bibfield  {title} {\bibinfo {title} {Reduced twinning and surface roughness
  of {Bi}$_2${Se}$_3$ and {Bi}$_2${Te}$_3$ layers grown by molecular beam
  epitaxy on sapphire substrates},\ }\href {https://doi.org/10.1116/1.5017977}
  {\bibfield  {journal} {\bibinfo  {journal} {Journal of Vacuum Science \&
  Technology B}\ }\textbf {\bibinfo {volume} {36}},\ \bibinfo {pages} {02D107}
  (\bibinfo {year} {2018})},\ \Eprint
  {https://arxiv.org/abs/https://doi.org/10.1116/1.5017977}
  {https://doi.org/10.1116/1.5017977} \BibitemShut {NoStop}%
\bibitem [{\citenamefont {Aramberri}\ \emph {et~al.}(2015)\citenamefont
  {Aramberri}, \citenamefont {Cerd{\'a}},\ and\ \citenamefont
  {Mu{\~{n}}oz}}]{r15_Aramberri_TPaDiracSt}%
  \BibitemOpen
  \bibfield  {author} {\bibinfo {author} {\bibfnamefont {H.}~\bibnamefont
  {Aramberri}}, \bibinfo {author} {\bibfnamefont {J.~I.}\ \bibnamefont
  {Cerd{\'a}}},\ and\ \bibinfo {author} {\bibfnamefont {M.~C.}\ \bibnamefont
  {Mu{\~{n}}oz}},\ }\bibfield  {title} {\bibinfo {title} {Tunable dirac
  electron and hole self-doping of topological insulators induced by stacking
  defects},\ }\href {https://doi.org/10.1021/acs.nanolett.5b00625} {\bibfield
  {journal} {\bibinfo  {journal} {Nano Letters}\ }\textbf {\bibinfo {volume}
  {15}},\ \bibinfo {pages} {3840} (\bibinfo {year} {2015})}\BibitemShut
  {NoStop}%
\bibitem [{\citenamefont {Skriver}(2012)}]{book_Skriver}%
  \BibitemOpen
  \bibfield  {author} {\bibinfo {author} {\bibfnamefont {H.~L.}\ \bibnamefont
  {Skriver}},\ }\href {https://doi.org/10.1007/978-3-642-81844-8} {\emph
  {\bibinfo {title} {The LMTO Method: Muffin-Tin Orbitals and Electronic
  Structure}}}\ (\bibinfo  {publisher} {Springer, Berlin},\ \bibinfo {year}
  {2012})\BibitemShut {NoStop}%
\bibitem [{\citenamefont {Turek}\ \emph {et~al.}(1997)\citenamefont {Turek},
  \citenamefont {Drchal}, \citenamefont {Kudrnovsky}, \citenamefont {Sob},\
  and\ \citenamefont {Weinberger}}]{book_Turek}%
  \BibitemOpen
  \bibfield  {author} {\bibinfo {author} {\bibfnamefont {I.}~\bibnamefont
  {Turek}}, \bibinfo {author} {\bibfnamefont {V.}~\bibnamefont {Drchal}},
  \bibinfo {author} {\bibfnamefont {J.}~\bibnamefont {Kudrnovsky}}, \bibinfo
  {author} {\bibfnamefont {M.}~\bibnamefont {Sob}},\ and\ \bibinfo {author}
  {\bibfnamefont {P.}~\bibnamefont {Weinberger}},\ }\href
  {https://doi.org/10.1007/978-1-4615-6255-9} {\emph {\bibinfo {title}
  {Electronic Structure of Disordered Alloys, Surfaces and Interfaces}}}\
  (\bibinfo  {publisher} {Kluwer, Boston},\ \bibinfo {year} {1997})\BibitemShut
  {NoStop}%
\bibitem [{\citenamefont {Vosko}\ \emph {et~al.}(1980)\citenamefont {Vosko},
  \citenamefont {Wilk},\ and\ \citenamefont {Nusair}}]{r80_Vosko_VWNpot}%
  \BibitemOpen
  \bibfield  {author} {\bibinfo {author} {\bibfnamefont {S.~H.}\ \bibnamefont
  {Vosko}}, \bibinfo {author} {\bibfnamefont {L.}~\bibnamefont {Wilk}},\ and\
  \bibinfo {author} {\bibfnamefont {M.}~\bibnamefont {Nusair}},\ }\bibfield
  {title} {\bibinfo {title} {Accurate spin-dependent electron liquid
  correlation energies for local spin density calculations: a critical
  analysis},\ }\href {https://doi.org/10.1139/p80-159} {\bibfield  {journal}
  {\bibinfo  {journal} {Canadian Journal of Physics}\ }\textbf {\bibinfo
  {volume} {58}},\ \bibinfo {pages} {1200} (\bibinfo {year} {1980})},\ \Eprint
  {https://arxiv.org/abs/https://doi.org/10.1139/p80-159}
  {https://doi.org/10.1139/p80-159} \BibitemShut {NoStop}%
\bibitem [{\citenamefont {Korzhavyi}\ \emph {et~al.}(1995)\citenamefont
  {Korzhavyi}, \citenamefont {Ruban}, \citenamefont {Abrikosov},\ and\
  \citenamefont {Skriver}}]{r95_Ruban_screen}%
  \BibitemOpen
  \bibfield  {author} {\bibinfo {author} {\bibfnamefont {P.~A.}\ \bibnamefont
  {Korzhavyi}}, \bibinfo {author} {\bibfnamefont {A.~V.}\ \bibnamefont
  {Ruban}}, \bibinfo {author} {\bibfnamefont {I.~A.}\ \bibnamefont
  {Abrikosov}},\ and\ \bibinfo {author} {\bibfnamefont {H.~L.}\ \bibnamefont
  {Skriver}},\ }\bibfield  {title} {\bibinfo {title} {Madelung energy for
  random metallic alloys in the coherent potential approximation},\ }\href
  {https://doi.org/10.1103/PhysRevB.51.5773} {\bibfield  {journal} {\bibinfo
  {journal} {Phys. Rev. B}\ }\textbf {\bibinfo {volume} {51}},\ \bibinfo
  {pages} {5773} (\bibinfo {year} {1995})}\BibitemShut {NoStop}%
\bibitem [{\citenamefont {Velick\'y}\ \emph {et~al.}(1968)\citenamefont
  {Velick\'y}, \citenamefont {Kirkpatrick},\ and\ \citenamefont
  {Ehrenreich}}]{r68_Velicky_cpa}%
  \BibitemOpen
  \bibfield  {author} {\bibinfo {author} {\bibfnamefont {B.}~\bibnamefont
  {Velick\'y}}, \bibinfo {author} {\bibfnamefont {S.}~\bibnamefont
  {Kirkpatrick}},\ and\ \bibinfo {author} {\bibfnamefont {H.}~\bibnamefont
  {Ehrenreich}},\ }\bibfield  {title} {\bibinfo {title} {Single-site
  approximations in the electronic theory of simple binary alloys},\ }\href
  {https://doi.org/10.1103/PhysRev.175.747} {\bibfield  {journal} {\bibinfo
  {journal} {Phys. Rev.}\ }\textbf {\bibinfo {volume} {175}},\ \bibinfo {pages}
  {747} (\bibinfo {year} {1968})}\BibitemShut {NoStop}%
\bibitem [{\citenamefont {Kudrnovsk\'y}\ \emph {et~al.}(2000)\citenamefont
  {Kudrnovsk\'y}, \citenamefont {Drchal}, \citenamefont {Blaas}, \citenamefont
  {Weinberger}, \citenamefont {Turek},\ and\ \citenamefont
  {Bruno}}]{r00_Kudrnovsky_layers}%
  \BibitemOpen
  \bibfield  {author} {\bibinfo {author} {\bibfnamefont {J.}~\bibnamefont
  {Kudrnovsk\'y}}, \bibinfo {author} {\bibfnamefont {V.}~\bibnamefont
  {Drchal}}, \bibinfo {author} {\bibfnamefont {C.}~\bibnamefont {Blaas}},
  \bibinfo {author} {\bibfnamefont {P.}~\bibnamefont {Weinberger}}, \bibinfo
  {author} {\bibfnamefont {I.}~\bibnamefont {Turek}},\ and\ \bibinfo {author}
  {\bibfnamefont {P.}~\bibnamefont {Bruno}},\ }\bibfield  {title} {\bibinfo
  {title} {Ab initio theory of perpendicular magnetotransport in metallic
  multilayers},\ }\href {https://doi.org/10.1103/PhysRevB.62.15084} {\bibfield
  {journal} {\bibinfo  {journal} {Phys. Rev. B}\ }\textbf {\bibinfo {volume}
  {62}},\ \bibinfo {pages} {15084} (\bibinfo {year} {2000})}\BibitemShut
  {NoStop}%
\bibitem [{\citenamefont {Turek}\ \emph {et~al.}(1995)\citenamefont {Turek},
  \citenamefont {Kudrnovsk\'y}, \citenamefont {\ifmmode~\check{S}\else
  \v{S}\fi{}ob}, \citenamefont {Drchal},\ and\ \citenamefont
  {Weinberger}}]{r95_Turek_layers}%
  \BibitemOpen
  \bibfield  {author} {\bibinfo {author} {\bibfnamefont {I.}~\bibnamefont
  {Turek}}, \bibinfo {author} {\bibfnamefont {J.}~\bibnamefont {Kudrnovsk\'y}},
  \bibinfo {author} {\bibfnamefont {M.}~\bibnamefont {\ifmmode~\check{S}\else
  \v{S}\fi{}ob}}, \bibinfo {author} {\bibfnamefont {V.}~\bibnamefont
  {Drchal}},\ and\ \bibinfo {author} {\bibfnamefont {P.}~\bibnamefont
  {Weinberger}},\ }\bibfield  {title} {\bibinfo {title} {Ferromagnetism of
  imperfect ultrathin {Ru} and {Rh} films on a ag(001) substrate},\ }\href
  {https://doi.org/10.1103/PhysRevLett.74.2551} {\bibfield  {journal} {\bibinfo
   {journal} {Phys. Rev. Lett.}\ }\textbf {\bibinfo {volume} {74}},\ \bibinfo
  {pages} {2551} (\bibinfo {year} {1995})}\BibitemShut {NoStop}%
\bibitem [{\citenamefont {Kudrnovský}\ \emph {et~al.}(2002)\citenamefont
  {Kudrnovský}, \citenamefont {Drchal}, \citenamefont {Turek}, \citenamefont
  {Dederichs}, \citenamefont {Weinberger},\ and\ \citenamefont
  {Bruno}}]{r02_Kudrnovsky_layers}%
  \BibitemOpen
  \bibfield  {author} {\bibinfo {author} {\bibfnamefont {J.}~\bibnamefont
  {Kudrnovský}}, \bibinfo {author} {\bibfnamefont {V.}~\bibnamefont {Drchal}},
  \bibinfo {author} {\bibfnamefont {I.}~\bibnamefont {Turek}}, \bibinfo
  {author} {\bibfnamefont {P.}~\bibnamefont {Dederichs}}, \bibinfo {author}
  {\bibfnamefont {P.}~\bibnamefont {Weinberger}},\ and\ \bibinfo {author}
  {\bibfnamefont {P.}~\bibnamefont {Bruno}},\ }\bibfield  {title} {\bibinfo
  {title} {Ab initio theory of perpendicular transport in layered magnetic
  systems},\ }\href
  {https://doi.org/https://doi.org/10.1016/S0304-8853(01)00748-X} {\bibfield
  {journal} {\bibinfo  {journal} {Journal of Magnetism and Magnetic Materials}\
  }\textbf {\bibinfo {volume} {240}},\ \bibinfo {pages} {177 } (\bibinfo {year}
  {2002})},\ \bibinfo {note} {4th International Symposium on Metallic
  Multilayers}\BibitemShut {NoStop}%
\bibitem [{\citenamefont {Spedding}\ \emph {et~al.}(1956)\citenamefont
  {Spedding}, \citenamefont {Daane},\ and\ \citenamefont
  {Herrmann}}]{r56_Spedding_scandium}%
  \BibitemOpen
  \bibfield  {author} {\bibinfo {author} {\bibfnamefont {F.~H.}\ \bibnamefont
  {Spedding}}, \bibinfo {author} {\bibfnamefont {A.~H.}\ \bibnamefont
  {Daane}},\ and\ \bibinfo {author} {\bibfnamefont {K.~W.}\ \bibnamefont
  {Herrmann}},\ }\bibfield  {title} {\bibinfo {title} {{The crystal structures
  and lattice parameters of high-purity scandium, yttrium and the rare earth
  metals}},\ }\href {https://doi.org/10.1107/S0365110X5600156X} {\bibfield
  {journal} {\bibinfo  {journal} {Acta Crystallographica}\ }\textbf {\bibinfo
  {volume} {9}},\ \bibinfo {pages} {559} (\bibinfo {year} {1956})}\BibitemShut
  {NoStop}%
\bibitem [{\citenamefont {Vališka}\ \emph {et~al.}(2016)\citenamefont
  {Vališka}, \citenamefont {Warmuth}, \citenamefont {Michiardi}, \citenamefont
  {Vondráček}, \citenamefont {Ngankeu}, \citenamefont {Holý}, \citenamefont
  {Sechovský}, \citenamefont {Springholz}, \citenamefont {Bianchi},
  \citenamefont {Wiebe}, \citenamefont {Hofmann},\ and\ \citenamefont
  {Honolka}}]{r16_Valiska_BiSeHeterostructures}%
  \BibitemOpen
  \bibfield  {author} {\bibinfo {author} {\bibfnamefont {M.}~\bibnamefont
  {Vališka}}, \bibinfo {author} {\bibfnamefont {J.}~\bibnamefont {Warmuth}},
  \bibinfo {author} {\bibfnamefont {M.}~\bibnamefont {Michiardi}}, \bibinfo
  {author} {\bibfnamefont {M.}~\bibnamefont {Vondráček}}, \bibinfo {author}
  {\bibfnamefont {A.~S.}\ \bibnamefont {Ngankeu}}, \bibinfo {author}
  {\bibfnamefont {V.}~\bibnamefont {Holý}}, \bibinfo {author} {\bibfnamefont
  {V.}~\bibnamefont {Sechovský}}, \bibinfo {author} {\bibfnamefont
  {G.}~\bibnamefont {Springholz}}, \bibinfo {author} {\bibfnamefont
  {M.}~\bibnamefont {Bianchi}}, \bibinfo {author} {\bibfnamefont
  {J.}~\bibnamefont {Wiebe}}, \bibinfo {author} {\bibfnamefont
  {P.}~\bibnamefont {Hofmann}},\ and\ \bibinfo {author} {\bibfnamefont
  {J.}~\bibnamefont {Honolka}},\ }\bibfield  {title} {\bibinfo {title}
  {Topological insulator homojunctions including magnetic layers: The example
  of n-p type (n-qls bi$_2$se$_3$/mn-bi$_2$se$_3$) heterostructures},\ }\href
  {https://doi.org/10.1063/1.4954834} {\bibfield  {journal} {\bibinfo
  {journal} {Applied Physics Letters}\ }\textbf {\bibinfo {volume} {108}},\
  \bibinfo {pages} {262402} (\bibinfo {year} {2016})},\ \Eprint
  {https://arxiv.org/abs/https://doi.org/10.1063/1.4954834}
  {https://doi.org/10.1063/1.4954834} \BibitemShut {NoStop}%
\bibitem [{\citenamefont {Carva}\ \emph {et~al.}(2020)\citenamefont {Carva},
  \citenamefont {Bal\'a\ifmmode~\check{z}\else \v{z}\fi{}}, \citenamefont
  {\ifmmode~\check{S}\else \v{S}\fi{}ebesta}, \citenamefont {Turek},
  \citenamefont {Kudrnovsk\'y}, \citenamefont {M\'aca}, \citenamefont {Drchal},
  \citenamefont {Chico}, \citenamefont {Sechovsk\'y},\ and\ \citenamefont
  {Honolka}}]{r20_Carva_BiSe}%
  \BibitemOpen
  \bibfield  {author} {\bibinfo {author} {\bibfnamefont {K.}~\bibnamefont
  {Carva}}, \bibinfo {author} {\bibfnamefont {P.}~\bibnamefont
  {Bal\'a\ifmmode~\check{z}\else \v{z}\fi{}}}, \bibinfo {author} {\bibfnamefont
  {J.}~\bibnamefont {\ifmmode~\check{S}\else \v{S}\fi{}ebesta}}, \bibinfo
  {author} {\bibfnamefont {I.}~\bibnamefont {Turek}}, \bibinfo {author}
  {\bibfnamefont {J.}~\bibnamefont {Kudrnovsk\'y}}, \bibinfo {author}
  {\bibfnamefont {F.}~\bibnamefont {M\'aca}}, \bibinfo {author} {\bibfnamefont
  {V.}~\bibnamefont {Drchal}}, \bibinfo {author} {\bibfnamefont
  {J.}~\bibnamefont {Chico}}, \bibinfo {author} {\bibfnamefont
  {V.}~\bibnamefont {Sechovsk\'y}},\ and\ \bibinfo {author} {\bibfnamefont
  {J.}~\bibnamefont {Honolka}},\ }\bibfield  {title} {\bibinfo {title}
  {Magnetic properties of {Mn}-doped {Bi}$_{2}${Se}$_{3}$ topological
  insulators: Ab initio calculations},\ }\href
  {https://doi.org/10.1103/PhysRevB.101.054428} {\bibfield  {journal} {\bibinfo
   {journal} {Phys. Rev. B}\ }\textbf {\bibinfo {volume} {101}},\ \bibinfo
  {pages} {054428} (\bibinfo {year} {2020})}\BibitemShut {NoStop}%
\bibitem [{\citenamefont {Liechtenstein}\ \emph {et~al.}(1987)\citenamefont
  {Liechtenstein}, \citenamefont {Katsnelson}, \citenamefont {Antropov},\ and\
  \citenamefont {Gubanov}}]{r87_Liechtenstein_exchange}%
  \BibitemOpen
  \bibfield  {author} {\bibinfo {author} {\bibfnamefont {A.}~\bibnamefont
  {Liechtenstein}}, \bibinfo {author} {\bibfnamefont {M.}~\bibnamefont
  {Katsnelson}}, \bibinfo {author} {\bibfnamefont {V.}~\bibnamefont
  {Antropov}},\ and\ \bibinfo {author} {\bibfnamefont {V.}~\bibnamefont
  {Gubanov}},\ }\bibfield  {title} {\bibinfo {title} {Local spin density
  functional approach to the theory of exchange interactions in ferromagnetic
  metals and alloys},\ }\href
  {https://doi.org/https://doi.org/10.1016/0304-8853(87)90721-9} {\bibfield
  {journal} {\bibinfo  {journal} {Journal of Magnetism and Magnetic Materials}\
  }\textbf {\bibinfo {volume} {67}},\ \bibinfo {pages} {65 } (\bibinfo {year}
  {1987})}\BibitemShut {NoStop}%
\bibitem [{\citenamefont {Turek}\ \emph {et~al.}(2006)\citenamefont {Turek},
  \citenamefont {Kudrnovský}, \citenamefont {Drchal},\ and\ \citenamefont
  {Bruno}}]{r06_Turek_exchange}%
  \BibitemOpen
  \bibfield  {author} {\bibinfo {author} {\bibfnamefont {I.}~\bibnamefont
  {Turek}}, \bibinfo {author} {\bibfnamefont {J.}~\bibnamefont {Kudrnovský}},
  \bibinfo {author} {\bibfnamefont {V.}~\bibnamefont {Drchal}},\ and\ \bibinfo
  {author} {\bibfnamefont {P.}~\bibnamefont {Bruno}},\ }\bibfield  {title}
  {\bibinfo {title} {Exchange interactions, spin waves, and transition
  temperatures in itinerant magnets},\ }\href
  {https://doi.org/10.1080/14786430500504048} {\bibfield  {journal} {\bibinfo
  {journal} {Philosophical Magazine}\ }\textbf {\bibinfo {volume} {86}},\
  \bibinfo {pages} {1713} (\bibinfo {year} {2006})},\ \Eprint
  {https://arxiv.org/abs/https://doi.org/10.1080/14786430500504048}
  {https://doi.org/10.1080/14786430500504048} \BibitemShut {NoStop}%
\bibitem [{\citenamefont {Polyakov}\ \emph {et~al.}(2015)\citenamefont
  {Polyakov}, \citenamefont {Meyerheim}, \citenamefont {Crozier}, \citenamefont
  {Gordon}, \citenamefont {Mohseni}, \citenamefont {Roy}, \citenamefont
  {Ernst}, \citenamefont {Vergniory}, \citenamefont {Zubizarreta},
  \citenamefont {Otrokov}, \citenamefont {Chulkov},\ and\ \citenamefont
  {Kirschner}}]{r15_Polyakov_BiFeSe}%
  \BibitemOpen
  \bibfield  {author} {\bibinfo {author} {\bibfnamefont {A.}~\bibnamefont
  {Polyakov}}, \bibinfo {author} {\bibfnamefont {H.~L.}\ \bibnamefont
  {Meyerheim}}, \bibinfo {author} {\bibfnamefont {E.~D.}\ \bibnamefont
  {Crozier}}, \bibinfo {author} {\bibfnamefont {R.~A.}\ \bibnamefont {Gordon}},
  \bibinfo {author} {\bibfnamefont {K.}~\bibnamefont {Mohseni}}, \bibinfo
  {author} {\bibfnamefont {S.}~\bibnamefont {Roy}}, \bibinfo {author}
  {\bibfnamefont {A.}~\bibnamefont {Ernst}}, \bibinfo {author} {\bibfnamefont
  {M.~G.}\ \bibnamefont {Vergniory}}, \bibinfo {author} {\bibfnamefont
  {X.}~\bibnamefont {Zubizarreta}}, \bibinfo {author} {\bibfnamefont {M.~M.}\
  \bibnamefont {Otrokov}}, \bibinfo {author} {\bibfnamefont {E.~V.}\
  \bibnamefont {Chulkov}},\ and\ \bibinfo {author} {\bibfnamefont
  {J.}~\bibnamefont {Kirschner}},\ }\bibfield  {title} {\bibinfo {title}
  {Surface alloying and iron selenide formation in
  {Fe/Bi}$_{2}${Se}$_{3}(0001)$ observed by x-ray absorption fine structure
  experiments},\ }\href {https://doi.org/10.1103/PhysRevB.92.045423} {\bibfield
   {journal} {\bibinfo  {journal} {Phys. Rev. B}\ }\textbf {\bibinfo {volume}
  {92}},\ \bibinfo {pages} {045423} (\bibinfo {year} {2015})}\BibitemShut
  {NoStop}%
\end{thebibliography}%

\end{document}